\documentclass {agujournal2019}
\usepackage{latexsym}
\usepackage[utf8]{inputenc}
\usepackage{graphicx}
\usepackage{url}
\usepackage{ragged2e}
\usepackage{array}
\usepackage{soul}
\usepackage{subcaption}
\journalname{JGR: Space Physics}

\begin{document}

\title{Radial Evolution of Coronal Mass Ejections Between MESSENGER, \textit{Venus Express}, STEREO, and L1: Catalog and Analysis}

\affiliation{1}{Space Science Center and Department of Physics, University of New Hampshire, Durham, NH, USA.}

\authors{T.\ M.~Salman\affil{1}, R.\ M.~Winslow\affil{1},  N.\ Lugaz\affil{1}}

\correspondingauthor{T. M. Salman}{ts1090@wildcats.unh.edu}

\begin{keypoints}
\item Comprehensive catalog of 47 CMEs observed in conjunction between radially aligned spacecraft in the inner heliosphere. 
\item Conjunction events reveal that the sheath duration expands nearly linearly but independently of the initial CME speed.
\item Average decrease of the maximum magnetic field strength in the CME agrees with past results, but individual CMEs may deviate significantly from the average.
\end{keypoints}

\begin{abstract}

\justify

Our knowledge of the properties of Coronal Mass Ejections (CMEs) in the inner heliosphere is constrained by the relative lack of plasma observations between Sun and 1 AU. In this work, we present a comprehensive catalog of 47 CMEs measured \textit{in situ} measurements by two or more radially aligned spacecraft (MESSENGER, \textit{Venus Express}, STEREO, or \textit{Wind}/ACE). We estimate the CME impact speeds at Mercury and Venus using a drag-based model and present an average propagation profile of CMEs (speed and deceleration/acceleration) in the inner heliosphere. We find that CME deceleration continues past Mercury's orbit but most of the deceleration occurs between the Sun and Mercury. We examine the exponential decrease of the maximum magnetic field strength in the CME with heliocentric distance using two approaches: a modified statistical method and analysis from individual conjunction events. Findings from both the approaches are on average consistent with previous studies but show significant event-to-event variability. We also find the expansion of the CME sheath to be well fit by a linear function. However, we observe the average sheath duration and its increase to be fairly independent of the initial CME speed, contradicting commonly held knowledge that slower CMEs drive larger sheaths. We also present an analysis of the 3 November 2011 CME observed in longitudinal conjunction between MESSENGER, \textit{Venus Express}, and STEREO-B focusing on the expansion of the CME and its correlation with the exponential fall-off of the maximum magnetic field strength in the ejecta.

\end{abstract}

\section{Introduction} \label{sec:intro}

\justify

Coronal Mass Ejections (CMEs) are large-scale, episodic transient events in which large amounts (on average 10$^{15}$-10$^{16}$ g of plasma) of material from the solar atmosphere are ejected in the solar wind \cite{Hundhausen:1988}. CMEs remove the built-up magnetic flux and helicity over the solar magnetic cycle from the solar corona \cite{Chen:2017,Green:2018,Low:1996}. Most of the ejected material comes from the low corona, although cooler, denser material likely of chromospheric or photospheric origin is also sometimes involved \cite{Webb:2012}. CMEs are commonly associated with different forms of solar activity such as eruptive prominences and solar flares \cite{Gosling:1974}. CME ejected material in the solar wind is a crucial link between solar activity and disturbances in the heliosphere \cite{Kilpua:2017b,Liu:2010}.

\justify

CMEs are responsible for the most extreme space weather effects at Earth \cite{Baker:2008}. \citeA{Zhang:2007} found 87{\%} of the 88 intense geomagnetic storms (Dst$\leq$-100 nT) that occurred during 1996-2005 to be caused by CMEs. Long-duration (\textgreater 3 h), large and negative (\textless -10~nT) southward B$_{z}$ lead to intense geo-magnetic storms \cite{Gonzalez:1987}. The key reasons why CMEs drive the strongest geomagnetic storms are their enhanced speed and magnetic field inside the magnetic ejecta and the sheath regions, which can provide strong southward fields for several hours \cite{Farrugia:1997,Huttunen:2005,Zhang:2007}. As a result, reliable and actionable space weather forecasting relies on the ability to predict the magnetic field structure and longevity of a CME before arrival at Earth \cite{Kilpua:2019}, provided that the magnetic field direction does not change drastically during the remaining propagation time \cite{Winslow:2016}. As the geoeffectiveness of CMEs strongly depends on the magnetic field direction within them, present efforts are focused on inferring their general magnetic topology in real time from space-based solar observatories \cite <e.g.,>[]{Palmerio:2018,Salman:2018,Savani:2015}.

\justify 

\textit{In situ} measurements along with spaceborne and ground-based remote observations have improved our knowledge of the origins and development of CMEs \cite{Howard:2012,Lugaz:2011a}. Before the launch of STEREO \cite <Solar Terrestrial Relations Observatory:>[]{Kaiser:2005}, most modern-day CME observations were primarily constrained to two domains: near-Earth space borne observations extending out to 30 solar radius and \textit{in situ} measurements near L1 (first Lagrangian Point). With the launch of STEREO in 2006, CMEs have been routinely observed from the Sun to the Earth combining observations from extreme ultraviolet imager (EUVI: range of 1-1.7 R$_{s}$), coronagraphs (COR1: observes white light ranging from 1.5-4 R$_{s}$ and COR2: observes white light ranging from 2.5-15 R$_{s}$) and Heliospheric Imagers (HI-1: range of 15-84 R$_{s}$ and HI-2: 66-318 R$_{s}$) on board the STEREO spacecraft \cite{Davis:2009,Harrison:2005,Harrison:2018,Howard:2008}. However, there is often a mismatch between the inferred magnetic configuration of CMEs near the Sun as determined from various indirect proxies (coronagraph and heliospheric imagers do not give direct information about the magnetic field inside CMEs) and what is measured \textit{in situ} \cite <e.g.,>[]{Palmerio:2018}. Probable reasons for this mismatch can be: the flux rope orientation (and, hence magnetic field rotation) is difficult to determine directly from white-light observations \cite{Isavnin:2013,Thernisien:2009}, the magnetic structure can drastically change during the CME propagation (see next paragraph), and existing methods cannot link \textit{in situ} signatures to coronal observations with reasonable precision. On the other hand, \textit{in situ} solar wind measurements provide a detailed but extremely limited and potentially localized view of a CME at the position of the spacecraft near 1 AU \cite{Lugaz:2018}. Due to expansion, at this distance the CME can cover up to around 100$^{\circ}$ in heliocentric longitude and several tenths of an AU along the radial direction to the Sun \cite{Bothmer:1998,Liu:2005,Moestl:2012,Richardson:2010,Wood:2010}. \citeA{Good:2016} focused on the longitudinal extent of CMEs based on the proportion of CMEs measured by two spacecraft for various longitudinal separations. They found that when two spacecraft were within 30$^{\circ}$ from each other, signatures of the flux rope are observed by both spacecraft 65{\%} of the time. In addition, there is a lack of plasma and compositional data at sub-1 AU heliocentric distances as recent planetary missions in the inner heliosphere such as MESSENGER \cite <MErcury Surface, Space ENvironment, GEochemistry, and Ranging:>[]{Solomon:2001} and \textit{Venus Express} \cite{Titov:2006} do not typically provide solar wind plasma measurements. Therefore, we have a key observational gap in our understanding of CME evolution and propagation into interplanetary space between these domains \cite{Forsyth:2006}. However, with MAVEN \cite <Mars Atmosphere and Volatile Evolution:>[]{Jakosky:2015}, Parker Solar Probe \cite{Fox:2016}, and BepiColombo \cite{Benkhoff:2010} already in space and the future launch of Solar Orbiter \cite{Mueller:2013}, all missions including solar wind plasma measurements, we are approaching a new era of inner heliosphere exploration through dedicated missions.

\justify 

Coronal magnetic field and plasma carried by a CME in the heliosphere has distinctive signatures with a large amount of variation. However, except for a few studies that used multi-point spacecraft observations \cite{Burlaga:1981,Cane:1997,Farrugia:2011,Good:2015,Good:2018,Janvier:2019,Kilpua:2011,Moestl:2009b,Moestl:2009c,Moestl:2015,Prise:2015,Wang:2018,Winslow:2016,Winslow:2018}, CME measurements are largely restricted to single-point observations in space \cite <see also discussion in>[]{Lugaz:2018}. \citeA{Leitner:2007} previously listed 7 magnetic cloud (MC) events which were observed at two or more spacecraft during solar cycles 20 and 21. These lineup events were observed at heliocentric distances ranging from 0.62-9.4 AU, with longitudinal separations between the observing spacecraft being less than 20$^{\circ}$. Recently, \citeA{Good:2019} examined the self similarity in the time series profiles of magnetic field structures of 18 interplanetary flux ropes. These flux ropes were observed by radially aligned spacecraft in the inner heliosphere (combinations of MESSENGER, \textit{Venus Express}, and STEREO) with the latitudinal and longitudinal separations between the observing spacecraft not exceeding 15$^{\circ}$. 

\justify
 
Given this localized nature of CME observations, global configuration of a CME can be difficult to extrapolate as it is not always clear which part of the CME is being sampled, as a single spacecraft can only sample a narrow trajectory through the larger CME structure \cite{Reinard:2012}. Also the motion of a CME in the solar wind depends on the heliospheric environment it encounters during its propagation \cite{Kilpua:2012,Temmer:2011}. Therefore, CME signatures observed with near-Earth \textit{\textit{in situ}} measurements can drastically change from its coronal counterpart or other measurements in the inner heliosphere because of interaction with the background solar wind \cite <e.g. deceleration/acceleration, deformation, shock wave formation etc., see>[]{Manchester:2017}, deflection \cite <e.g.,>[]{Gosling:1987b,Isavnin:2013,Kay:2013,Kay:2015,Kilpua:2009,Lugaz:2011c,Vandas:1996,Wang:2004,Wang:2014,Zhuang:2019}, rotation \cite{Cohen:2010,Isavnin:2014,Nieves:2013}, interaction with another CME \cite{Lugaz:2015a,Lugaz:2017b}. With currently no plasma measurements at sub-1 AU heliocentric distances, we rely on models to simulate CME evolution in the inner heliosphere. Although present models based on numerical simulations \cite{Jin:2017,Lugaz:2007,Odstrcil:2002,Odstrcil:2004} or empirical methods \cite{Gopalswamy:2001b,Kay:2017,Kim:2007,Vrsnak:2012} can forecast the basic characteristics of CMEs (speed, arrival time), it cannot predict the exact magnetic structure of the CME and the development of the sheath region. As a result, it is important to have multi-point observations to understand the governing physics behind CME propagation and evolution in the inner heliosphere. That prompted us to search for conjunction events between radially aligned spacecraft in the inner heliosphere. Due to the orbital periods of Mercury (88 days) and Venus (225 days), the occurrence of radial conjunctions are frequent between these spacecraft and the ones at 1 AU (e.g. ACE, \textit{Wind}, STEREO). As a result, MESSENGER, \textit{Venus Express}, STEREO, and \textit{Wind}/ACE have the potential to offer unprecedented \textit{in situ} coverage of interplanetary space and radial measurements of CMEs. Such events can allow us to determine the radial evolution of CMEs with much better cohesiveness. From an observational perspective, detailed analysis of these events can provide a continuous picture of CME propagation in realistic background solar wind conditions and improve the accuracy of the CME magnetic structure prediction by linking solar observations with \textit{in situ} measurements. In this study, we combine two distinct issues/types of research: multi-spacecraft measurements of the same CME with the spacecraft at the same radial distance \cite{Burlaga:1981,Kilpua:2011,Moestl:2009b} and multi-spacecraft measurements of the same CME with the spacecraft at the same longitude \cite{Farrugia:2011,Good:2015,Good:2018,Moestl:2015,Prise:2015,Wang:2018,Winslow:2016,Winslow:2018}. Recently, \citeA{Janvier:2019} used data from MESSENGER, \textit{Venus Express}, and ACE to determine statistically how CME magnetic field, of both ejecta and sheath, evolve with radial distance \cite <see also>[]{Gulisano:2010,Winslow:2015}. Here, we take a different approach, where we focus on events where two or more spacecraft measure the same CME. By doing so, we do not have to rely on statistical techniques to determine how properties change with distance. We can also better evaluate how individual CME events have an evolution that deviate from the average statistical behavior of CMEs.

\justify

In this study, we list 45 CMEs observed in conjunction between at least two spacecraft (47 two-spacecraft conjunctions). Coronagraph observations and database timings are used in association to identify potential multi-spacecraft conjunction events. For each of these events, we require the longitudinal separation between the spacecraft to be less than 35$^{\circ}$ (except one, see Section~\ref{ssec:selection}). LASCO \cite <Large Angle Spectroscopic Coronagraph:>[]{Brueckner:1995} CMEs (with angular width \textgreater30$^{\circ}$) have an average angular width of 60$^{\circ}$ \cite{Gopalswamy:2010a}. Therefore a longitudinal separation less than 35$^{\circ}$ increases the likelihood of two or more radially aligned spacecraft observing both the sheath and the ejecta, rather than only the sheath itself \cite <see statistical results of>[]{Good:2016}. However, CMEs measured at longitudinal separations as small as 1$^{\circ}$ can have significant differences in measurements from one observing spacecraft to another \cite{Kilpua:2011,Lugaz:2018}. We discuss how this may affect the results later on in the text. 

\justify

The paper is organized as follows. In Section~\ref{sec:methodology}, we explain the method used to identify the coronal source of a CME counterpart observed at multi-spacecraft locations. We list 47 conjunction events between MESSENGER, \textit{Venus Express}, STEREO, and/or \textit{Wind}/ACE in Section~\ref{sec:database}. In Section~\ref{sec:case study}, we present an analysis of the 3 November 2011 CME event observed at three different points in space (MESSENGER, \textit{Venus Express}, STEREO-B). A brief summary and discussion are stated in Section~\ref{sec:conclusion}.

\section{Methodology} \label{sec:methodology}

\subsection{Description of the Data Set} \label{ssec:data}

\justify 

The MESSENGER spacecraft launched on 3 August 2004 was inserted into orbit about Mercury on 18 March 2011 following a 7-year cruise phase. MESSENGER is the first spacecraft since Helios 1 and 2 in the 1980s to make \textit{in situ} measurements of the interplanetary medium at heliocentric distances \textless 0.5 AU. The initial orbit had a 12 h period. MESSENGER remained in this orbit until 16 April 2012, when the apoapsis was decreased and the orbital period reduced to 8 h. During the initial orbit phase, MESSENGER typically spent 2-4 h per orbit inside Mercury's bow shock and magnetosphere and the rest of the time in the interplanetary medium. After the orbital period was lowered to 8 h, the time inside Mercury's magnetosphere increased to 3-5 h on average per orbit. The mission was terminated on 30 April 2015. In the present study, we have used 1-s high-resolution data from the magnetometer \cite <MAG:>[]{Anderson:2007} instrument on-board the MESSENGER spacecraft.

\justify 

The \textit{Venus Express} spacecraft was launched on 9 November 2005 to study the atmosphere of Venus, from the surface to the ionosphere. The spacecraft entered orbit on 11 April 2006 where it remained until the end of its mission in December 2014. In the present study, \textit{Venus Express's} MAG \cite{Zhang:2006} data with 1-min resolution is used.

\justify

STEREO, launched in 2006 employs two nearly identical Sun-pointed space-based observatories- one ahead of Earth in its orbit (STEREO-A), the other trailing behind (STEREO-B). Communications with the STEREO-B spacecraft were interrupted on 1 October 2014. In the present study, we have used 1-minute resolution plasma data from the PLASTIC instrument \cite{Galvin:2008} and 1/8 s high resolution magnetic field data from the IMPACT instrument \cite{Luhmann:2008} on-board the twin STEREO spacecraft.

\justify
 
The \textit{Wind} and Advanced Composition Explorer \cite <ACE:>[]{Stone:1998} were launched on 1 November 1994 and 25 August 1997 respectively and both now orbits the L1 point. In the present study, we have used 1-minute resolution data from the \textit{Wind} Magnetic Field Investigation \cite <MFI:>[]{Lepping:1995} instrument and 16-s high resolution data from the ACE Magnetic Fields Experiment instrument \cite{Smith:1998}.

\justify

We use the catalogs of CMEs observed by MESSENGER \cite <69 CMEs in the time interval 2011-2015:>{Winslow:2015,Winslow:2017}, \textit{Venus Express} \cite <84 events in the time interval 2006-2013:>{Good:2016}, STEREO \cite <2007-2016:>{Jian:2018}, and ACE \cite <1996-2018:>{Richardson:2010} to build our conjunction database of multi-point CME observations.

\subsection{Identification of Probable CME Candidate/s for a Conjunction Event} \label{ssec:identification}

\justify

For a CME to be listed as a possible conjunction between two or more spacecraft, we first require the longitudinal separation between the spacecraft to be less than 35$^{\circ}$ in Heliographic Inertial (HGI) Coordinates during the event. We start by examining the CMEs observed by the MESSENGER Magnetometer between 2011 and 2015. For each CME observed by MESSENGER during this time span, we list the longitudinal separations between MESSENGER and \textit{Venus Express}, STEREO, and \textit{Wind}/ACE. For any longitudinal separation greater than 35$^{\circ}$, we remove that specific event from consideration. Then we search for CME signatures measured by \textit{Venus Express}, STEREO, and \textit{Wind}/ACE spacecraft in their corresponding databases within an expected interval after the CME was initially measured at MESSENGER. Any CME satisfying this criterion is listed as a possible conjunction between the spacecraft. We repeat the same procedure using the \textit{Venus Express} CME catalog spanning 7.5 years (2006-2013) to list possible conjunctions between \textit{Venus Express} and STEREO and \textit{Venus Express} and \textit{Wind}/ACE. Therefore, the conjunction database inherently splits into three sections: events observed in conjunction by MESSENGER and \textit{Venus Express} between 2011 and 2013, MESSENGER and STEREO or \textit{Wind} or ACE between 2011 and 2015, and \textit{Venus Express} and STEREO or \textit{Wind} or ACE between 2006 and 2013.

\justify      

For each conjunction event, we identify the date and time of the CME's first appearance in the LASCO/C2 field of view. As we lack plasma measurements of the CMEs encountered by MESSENGER and \textit{Venus Express}, with the help of the Drag-Based model (DBM) formulated by \citeA{Vrsnak:2012}, we provide estimated impact speeds at Mercury and Venus. As an input for this estimate, we use the ambient solar wind speed measured near 1 AU. For estimation of impact speeds at Mercury and Venus for any MESSENGER-\textit{Venus Express} conjunction, we use 392 km\,s$^{-1}$ as the solar wind speed (average upstream speed of 113 CME-driven shocks measured by the STEREO spacecraft from 2007-2016) as an input for the DBM. To determine the date and time of the likely CME's first C2 appearance, we use the CDAW (Coordinated Data Analysis Workshops) CME catalog \cite{Yashiro:2004} and SECCHI \cite <Sun Earth Connection Coronal and Heliospheric Investigation:>[]{Howard:2008} CME lists \cite{Robbrecht:2009}, automatically generated by CACTus (Computer Aided CME Tracking software). To identify a potential CME candidate, we search for agreement between the CME launch direction and positioning of the corresponding spacecraft during the event interval. We approximate the CME propagation direction (halo, west/east limb, and backsided) using three different field of views: the CME observed from SOHO (LASCO observations) and the two STEREO spacecraft (COR2 observations). This is achieved by comparing the relative heliographic longitudes of the spacecraft on the day of interest. We perform this for each CME upto a period of 3-5 days (3 if spacecraft 1 is MESSENGER and 5 if spacecraft 1 is \textit{Venus Express}) before the CME was observed at spacecraft 1. After this initial process of elimination, we use the DBM to estimate the CME arrival time at the spacecraft in consideration and match them with the timings listed in their corresponding catalogs. If the arrival time predicted by the DBM is in reasonable agreement with the listed timing, we list the CME as a possible candidate for the conjunction event. 

\justify

For a two spacecraft conjunction, spacecraft 1 (MESSENGER or \textit{Venus Express}) represents the first spacecraft to observe the signatures of the CME during its propagation and spacecraft 2 (\textit{Venus Express} or STEREO or \textit{Wind}/ACE) represents the second spacecraft to observe signatures of the same CME later on. In the case of a three spacecraft conjunction event, we have a third spacecraft where the same CME which was observed at spacecraft 1 and 2 is also observed.

\justify

In our catalog, we have 18 events for which the longitudinal separations between the observing spacecraft are 0-9.9$^{\circ}$, 14 events for which the longitudinal separations are 10-19.9$^{\circ}$, 9 events for which the longitudinal separations are 20-29.9$^{\circ}$, and 6 events for which the longitudinal separations are  30-44.2$^{\circ}$. The average longitudinal separation of our catalog events is 16.2$^{\circ}$. The 6 events for which the longitudinal separations are \textgreater30$^{\circ}$, two of them have longitudinal separations \textgreater35$^{\circ}$ (37.6$^{\circ}$ and 44.2$^{\circ}$) and they are actually the same CME. The reason for including this event with such a large longitudinal separation is explained in Section~\ref{ssec:selection}. Table-S1 of the Supporting Information lists our full database of conjunction events. Further details about this database is given later on in the text (see Section~\ref{ssec:db}).

\subsection{Estimation of Impact Speeds at Mercury and Venus} \label{ssec:estimation}

\justify

We calculate the average transit speeds between the Sun and spacecraft 1 (V$_{S-SC1}$), spacecraft 1 and spacecraft 2 (V$_{SC1-SC2}$), and the Sun and spacecraft 2 (V$_{S-SC2}$) using the CME take-off time at the Sun and listed arrival times at spacecraft 1 and 2. For consistency, we use the shock/discontinuity arrival times if available at all spacecraft for the calculation of average transit speeds. We use the magnetic ejecta arrival time if shock arrival time is not available at one of the spacecraft in consideration (e.g. due to the spacecraft being inside the magnetosphere). In 12 out of 47 cases, we observe disagreement (V$_{SC1-SC2}$ $>$ V$_{S-SC1}$) in the expected average transit speeds. This discrepancy could be due to the CME still accelerating in the low corona. In such cases, we use the CME time at 20 R$_{s}$ rather than the CME take-off time at the Sun to calculate the average transit speeds. However, using the CME time at 20 R$_{s}$ only fixed 2 out of the 12 discrepancies. Such trends are not entirely unexpected from slow CMEs, where the CMEs actually go through acceleration far out from the Sun. Checking the CME counterparts for these 12 events, we do indeed observe that most of these discrepancies (10 out of 12) are features of slow CMEs (initial speeds less than 700 km\,s$^{-1}$).

\justify

We use the DBM for an estimate of impact speeds at Mercury and Venus. The DBM relies on the assumption that the driving Lorentz force, which launches a CME, ceases in the upper corona and after this propagation distance, the sole dominant force governing the CME propagation is the magnetohydrodynamical equivalent of the aerodynamic drag. This model treats CMEs as single expanding bodies, propagating through an isotropic environment to which Newton's second law is applied. This model provides analytical solutions of the equation of motion where the drag acceleration has a quadratic dependence on the CME relative speed. Under the assumptions of constant drag coefficient and constant solar wind flow, with no CME-CME interaction, this model provides predictions for the arrival of the front boundary of the ejecta and its impact speed. However, the DBM has certain intrinsic drawbacks which can result in significant amount of uncertainties for events in which the CME still considerably accelerates beyond 20 R$_{s}$, the ambient solar-wind regime does not remain constant throughout the CME propagation, when there is CME-CME interaction, etc. \cite{Vrsnak:2001,Vrsnak:2010,Vrsnak:2012,Zic:2015}

\justify

For a given set of seven input parameters (CME start date, CME start time, starting radial distance of the CME, speed of the CME at the starting radial distance, drag parameter with unit of $10^{-7}$ km$^{-1}$, asymptotic solar wind speed, and target heliocentric distance), the DBM provides the impact speed for any target in the heliosphere. We use 20 R$_{s}$ as the starting radial distance (to minimize the effects of Lorentz force) and the 2nd-order speed at 20 R$_{s}$ (second-order polynomial fit to the height-time measurements evaluated when the CME is at a height of 20 R$_{s}$) listed in the CDAW catalog as the CME speed. For the asymptotic solar wind speed, we use the solar wind speed from \textit{in situ} measurements at STEREO or \textit{Wind} or ACE before they encounter the CME. For MESSENGER and STEREO or L1 and \textit{Venus Express} and STEREO or L1 conjunctions, we perform a three-step measurement process to get estimated impact speeds at Mercury and Venus using the DBM. In the first step, we change the only variable (drag parameter) in the DBM to match the CME arrival time at STEREO or L1. Then, we use that same drag parameter to get an estimated impact speed at Mercury or Venus. In the second step, as we have \textit{in situ} plasma measurements available at STEREO and L1, we change the drag parameter to match the maximum CME speed measured at STEREO or L1. After that, we use that drag parameter to get the second estimated impact speed at Mercury or Venus. The third estimated impact speed at Mercury or Venus is obtained by matching the listed CME arrival time at MESSENGER or \textit{Venus Express} using the DBM. These three estimated impact speeds are then averaged. It is important to highlight here that with large longitudinal separations, there is a significant probability of different parts of the ejecta/sheath impacting different observing spacecraft at different times, related to the radius of curvature of the CME. This will be specifically discussed for our case study (in Section~\ref{ssec:sheath} and Section~\ref{ssec:MC}). Without knowing the true CME shape, it is impossible to fully estimate the errors of these estimates. In Table-S2 of the Supporting Information, we list the three speed estimates obtained from the three constraints (arrival time at SC1, arrival time at SC2, arrival speed at SC2) as well as the input parameters used in the DBM calculation (drag parameter, initial CME speed, solar wind speed). These three estimates give us an estimate of the error of the procedure, listed as the standard deviation between the various speed estimates in Table-S2.

\justify

For any conjunction between only MESSENGER and \textit{Venus Express}, we adopt a slightly modified technique since there are no \textit{in situ} measurements available at both the spacecraft. In the first step, we try to match the listed CME arrival time at MESSENGER by changing the drag parameter in the DBM. It provides an estimation of the impact speed at Mercury and using the same drag parameter, we get an estimated impact speed at Venus. In the second step, we simply reverse the technique by using the DBM to match the listed CME arrival time at \textit{Venus Express}.     

\subsection{Conjunction Events with Three Point Observations} \label{ssec:selection}

\justify

We have two events with three point observations (MESSENGER, \textit{Venus Express}, and STEREO) in our catalog which are the most promising for studies of CME evolution in the inner heliosphere. The first event appears three times in our catalog: as a MESSENGER-\textit{Venus Express} conjunction event (1-2011, number tags explained in Section~\ref{ssec:db}), a MESSENGER-STEREO A conjunction event (6-2011), and a \textit{Venus Express}-STEREO A conjunction event (34-2011). The longitudinal separation between Mercury and Venus was 6.6$^{\circ}$, Mercury and STEREO-A was 44.2$^{\circ}$, and Venus and STEREO-A was 37.6$^{\circ}$. The heliocentric distances of MESSENGER, \textit{Venus Express}, and STEREO-A were 0.32 AU, 0.72 AU, and 0.96 AU respectively at the CME onset time. Though the longitudinal separation between Mercury-STEREO A and Venus-STEREO A exceeds our predefined criterion, the CME in question is a really fast halo CME with substantial angular extent and non-linear speed at 20 R$_{s}$ approaching 2281 km\,s$^{-1}$ (listed in the CDAW catalog). Therefore, we approximate this CME to be observerd at the three spacecraft in consideration.  

\justify

The second event also appears three times in our catalog: as a MESSENGER-\textit{Venus Express} conjunction event (3-2011), a MESSENGER-STEREO B conjunction event (8-2011) and a \textit{Venus Express}-STEREO B conjunction event (35-2011). The longitudinal separation between Mercury and Venus was 23.1$^{\circ}$, Mercury and STEREO-B was -4.8$^{\circ}$, and Venus and STEREO-B was -27.2$^{\circ}$. The heliocentric distances of MESSENGER, \textit{Venus Express}, and STEREO-B were 0.44 AU, 0.73 AU, and 1.09 AU respectively at the CME onset time. Analysis of the second conjunction event is presented in Section~\ref{sec:case study}. Previouly, radial evolution of this event has been extensively studied by \citeA{Good:2015, Good:2018}.

\section{Catalog of Conjunction Events with Multi-Spacecraft Observations} \label{sec:database}

\subsection{Database and List of Parameters} \label{ssec:db}

\justify

In Table-S1 of the Supporting Information, we list our full database. We first list the onset date and time of the probable CME candidate. If the CME was observed by LASCO, we report the average CME onset time as calculated in the CDAW catalog (average between first-order-constant speed and second-order-constant acceleration onset times). Otherwise, we report the time of the first STEREO/COR image containing the CME. We then list the arrival times of the shock/discontinuity, magnetic ejecta leading edge and trailing edge at spacecraft 1 and 2. Arrival times at MESSENGER are listed from \citeA{Winslow:2015}, \textit{Venus Express} from \citeA{Good:2016}, STEREO from \citeA{Jian:2018}, and L1 from \citeA{Richardson:2010}. We place a number tag on each conjunction event (e.g. 1-2011, 2-2011 etc.,) to recall it if necessary.

\justify

We also list the heliocentric distances of the spacecraft at the CME onset time, longitudinal separation between the spacecraft when the discontinuity/ejecta arrives at spacecraft 1, and the maximum magnetic field strength observed in the CME (including both the sheath and the ejecta) at each spacecraft in Table-S1. The heliocentric distances and longitudinal separations [in HGI coordinates] are listed from HelioWeb (\url{https://cohoweb.gsfc.nasa.gov/coho/helios/heli.html}). In Table-S1, we also list the initial speed of the CME. For the speed, we select the coronagraph which observed the CME closest to a limb event. When LASCO observed the CME as a limb event, we report the second-order CME speed at 20 R$_{s}$ listed in the CDAW catalog. For STEREO observations, we report the maximum speed as listed in the CACTus catalog. 

\justify

In Table-S2 of the Supporting Information, we provide detailed information regarding our three-step DBM procedure. We list the drag parameters associated with each of the three estimates (two estimates for a MESSENGER-\textit{Venus Express} conjunction), the averaged impact speeds at Mercury and Venus, standard deviation of these estimates, solar wind speed and the maximum CME speed (when spacecraft 2 is STEREO/\textit{Wind}/ACE) from \textit{in situ} measurements near Earth. We also list the average transit speeds between the Sun and spacecraft 1 (V$_{S-SC1}$), spacecraft 1 and spacecraft 2 (V$_{SC1-SC2}$), and the Sun and spacecraft 2 (V$_{S-SC2}$).

\justify

In most of the cases, we are able to find one CME candidate in coronagraphic observations which fits the temporal and directional requirements of a specific conjunction event. If there is more than one potential CME candidate, we aim to isolate one candidate based on the estimated arrival time at the corresponding spacecraft using the DBM. We understand that there is some uncertainty in determining the CME counterpart based on arrival time agreements. As a result, if there are more than one CME candidate for a single conjunction event and all of them appear to be suitable matches, we list them all (conjunction events with number tags 31-2011, 35-2011, 37-2011, 40-2012). For the 31-2011 event, we highlight two potential CME counterparts on 9 April 2011 with LASCO onset times of 15:48 UT and 18:00 UT. We believe the consequent CME signatures observed at \textit{Venus Express} and STEREO-A are either due to an interaction between these two CMEs or the first one is the more suitable candidate. There are, however, two events in our CME catalog that do not have a well-defined CME counterpart. The first one is event: 30-2011 for which we did not identify a corresponding CME at the Sun (due to LASCO data gap). We have three events bearing number tags 28-2010 for which we are not confident in their CME counterparts. These events correspond to three CME measurements at \textit{Venus Express} within a day (1 August 2010 - 2 August 2010). The longitudinal separations between \textit{Venus Express} and STEREO-B was 16.5$^{\circ}$ on 1 August 2010 and 17.3$^{\circ}$ on 2 August 2010 in HGI coordinates. However, in the expected arrival interval at STEREO-B, only two CMEs were measured (both of these CMEs did not drive shocks and had no sheath regions). Therefore, we have two potential conjunction events. We list the CMEs observed at 3:39 UT and 8:24 UT on 1 August 2010 as the suitable CME counterparts. However, without the availavility of plasma measurements at \textit{Venus Express}, we were not able to link in a unique manner the \textit{Venus Express} and STEREO-B events with a particular CME counterpart at the Sun.

\subsection{Speed Profile} \label{ssec:profile}

\justify

\begin{figure*}[htbp!]
  \centering
        \includegraphics[width=1.0\linewidth]{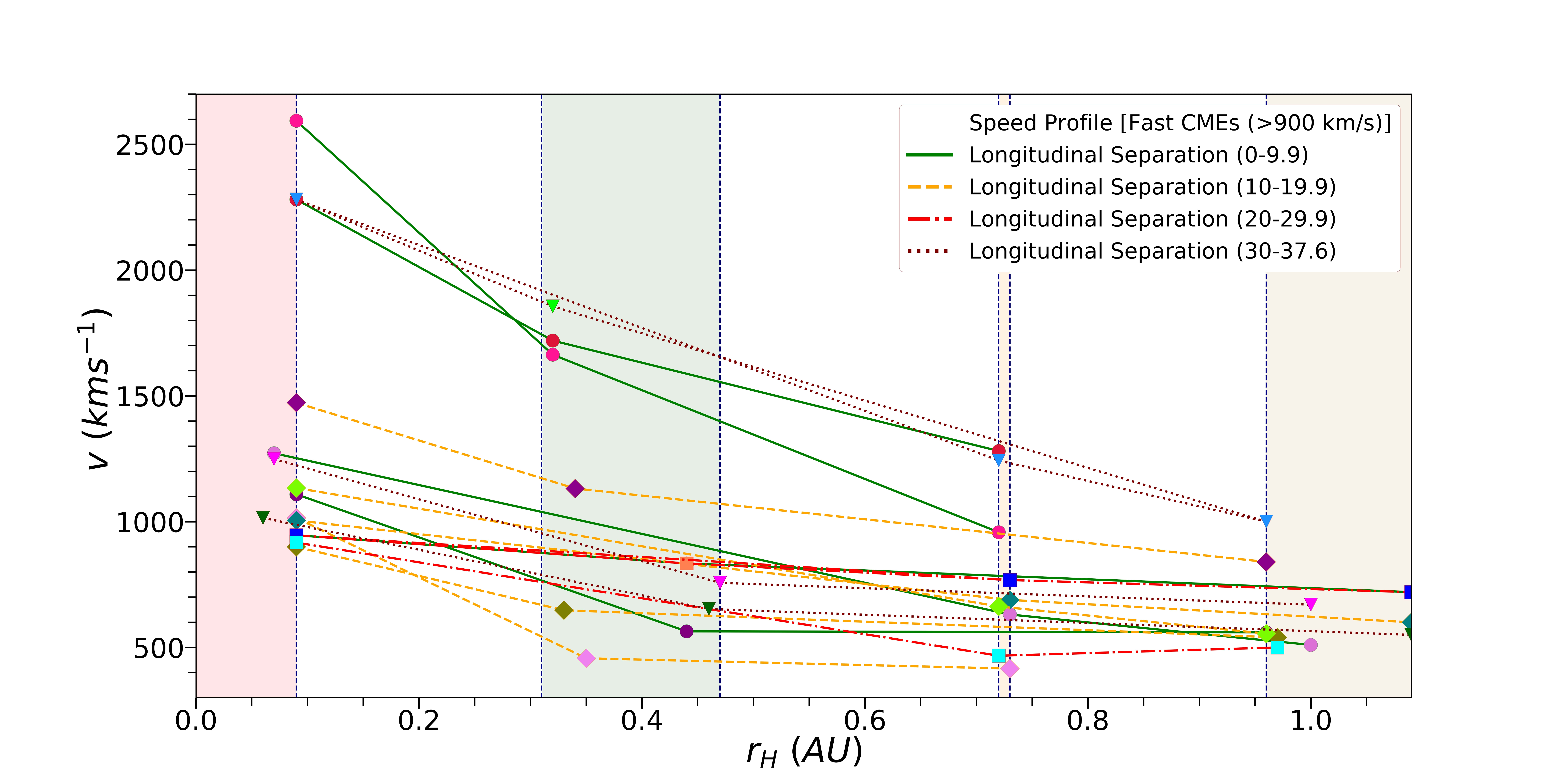}
        \caption{Propagation of fast CMEs with heliocentric distance (in AU). Lines connect measurements of the same CME and are color-coded depending on the longitudinal separation between the two \textit{in situ} spacecraft. Estimates of the errors for the speeds are discussed in the text.}
         \label{fig:fast}
  \end{figure*}

\begin{figure*}[htbp!]
  \centering
        \includegraphics[width=1.0\linewidth]{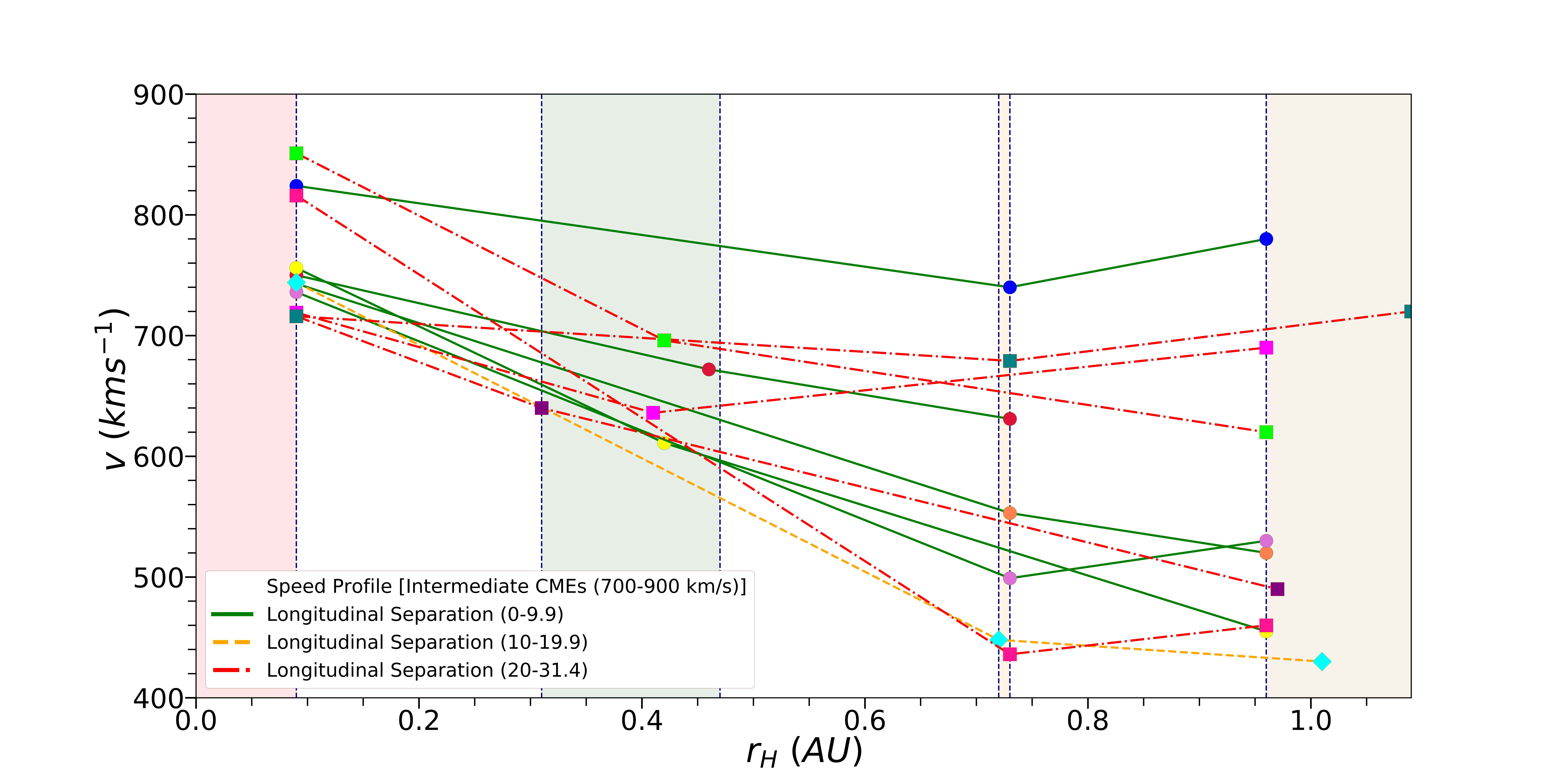}
        \caption{Same as Figure~\ref{fig:fast} but for CMEs of intermediate speed.}
         \label{fig:int}
  \end{figure*}

\begin{figure*}[htbp!]
  \centering
        \includegraphics[width=1.0\linewidth]{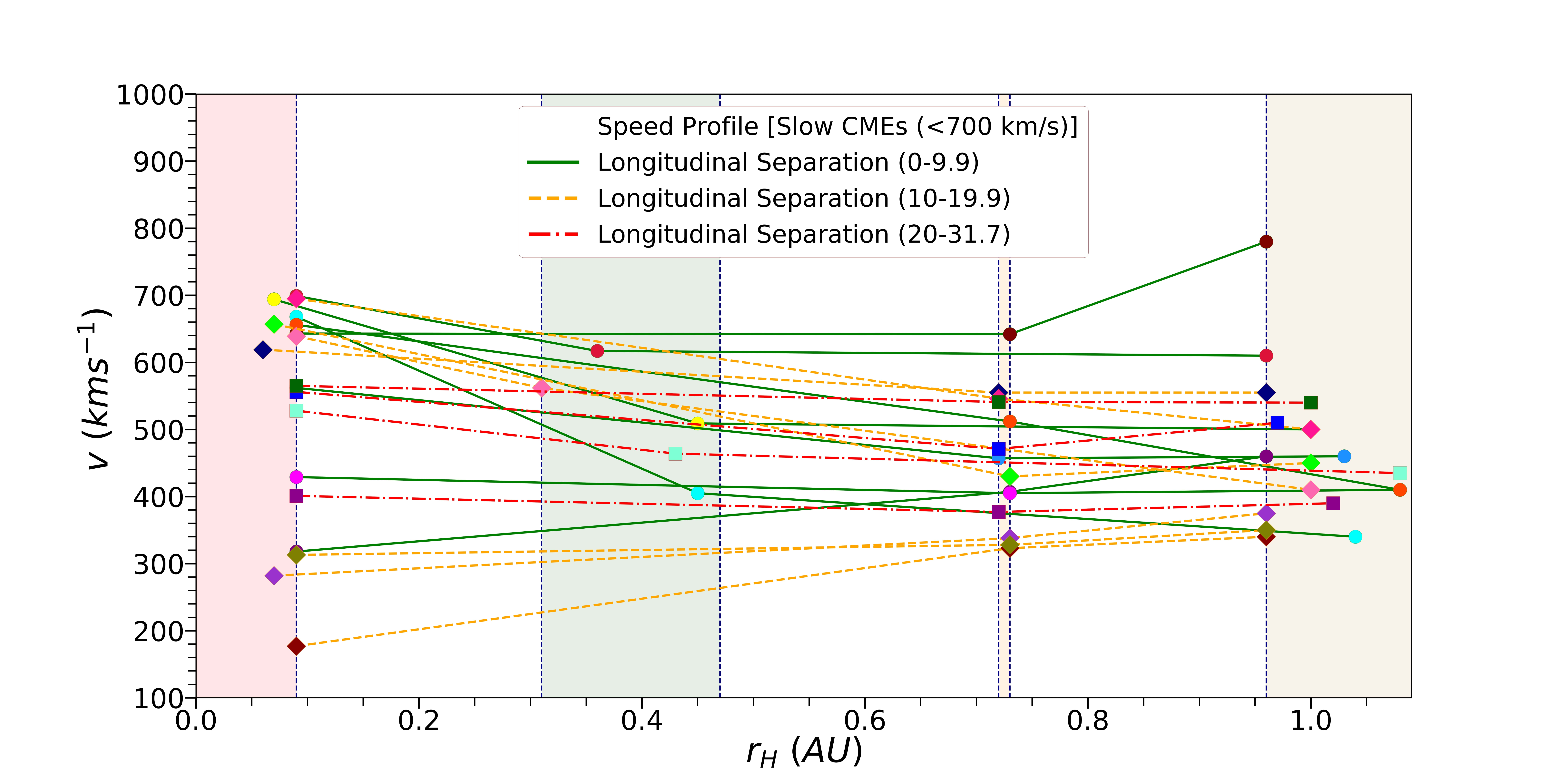}
        \caption{Same as Figure~\ref{fig:fast} but for slow CMEs.}
         \label{fig:slow}
  \end{figure*}

The basic characteristics of CME propagation is: CMEs that are faster than the ambient solar wind are decelerated, whereas those slower than the solar wind are accelerated by the ambient flow \cite{Gopalswamy:2000,Lindsay:1999}. Figure~\ref{fig:fast}, Figure~\ref{fig:int}, and Figure~\ref{fig:slow} show the propagation speed profiles of the CMEs from the Sun to spacecraft 2. The initial speed mentioned here is either the second-order CME speed at 20 R$_{s}$ listed in the CDAW catalog or the maximum speed listed in the CACTus catalog. For MESSENGER-STEREO/L1 conjunction events, we have the initial CME speed, average estimated impact speed at Mercury from the DBM, and the maximum CME speed measured at STEREO/\textit{Wind}/ACE. For \textit{Venus Express}-STEREO/L1 conjunction events, we have the initial CME speed, average estimated impact speed at Venus using the DBM, and the maximum CME speed measured at STEREO/\textit{Wind}/ACE. For MESSENGER-\textit{Venus Express} conjunction events, we have the initial CME speed and average estimated impact speeds at Mercury and Venus from the DBM. For coronagraphic observations, we use the speed obtained from the best observing spacecraft, for which the CME is close to a limb event. Limb views significantly minimize projection effects as compared to halo views and provide a better estimate of CME speeds. While we chose the best-observing spacecraft to minimize any projection effects, using speeds obtained from catalogs come with large error bars. Typical uncertainty range for these catalogs have been reported to be of the order of $\sim$150-200 km\,s$^{-1}$, both for individual events and for the difference between the two catalogs themselves \cite <see>[]{Michalek:2017,Robbrecht:2009}. Error associated with each DBM estimate is listed in Table-S2 of the Supporting Information. We also cannot rule out the possibility of different parts of the ejecta being sampled at different observing spacecraft due to a large longitudinal extent. To further highlight the importance of longitudinal separations, in the propagation speed profiles, we bin the CMEs based on the longitudinal separations between the spacecraft observing the conjunction event to qualitatively represent potential scenarios concerning higher uncertainties due to large longitudinal separations between two observation points. The four shaded regions in Figure~\ref{fig:fast}, Figure~\ref{fig:int}, and Figure~\ref{fig:slow} represent the following domains (from left to right): i) region closer to the Sun extending out to 20 R$_{s}$, ii) region of MESSENGER measurements (0.31-0.44 AU), iii) region of \textit{Venus Express} measurements (0.72-0.73 AU), iv) region of \textit{in situ} measeurements near 1 AU with STEREO and \textit{Wind}/ACE (0.96-1.09 AU). Each line connecting the three individual CME speeds at different heliocentric domains represents the propagation profile of the same CME in the inner heliosphere after its eruption and are color-coded depending on the longitudinal separation between the two measuring \textit{in situ} spacecraft. The CMEs are separated into three categories based on their initial speeds as measured by coronagraphs: fast (initial speeds greater than 900 km\,s$^{-1}$, see Figure~\ref{fig:fast}), slow (initial speeds less than 700 km\,s$^{-1}$, see Figure~\ref{fig:slow}) and intermediate (initial speeds between 700-900 km\,s$^{-1}$, see Figure~\ref{fig:int}).

\justify

For fast CMEs, we observe on average a 45{\%} decrease in speeds through their propagation from the Sun to spacecraft 2 (\textit{Venus Express} or STEREO or \textit{Wind}/ACE). Most of the variation (30{\%}) occurs in the innermost heliosphere, within Mercury's orbit, while the rest (15{\%}) occurs past 0.44 AU. For intermediate-speed CMEs, this average decrease in speeds drops to 24{\%} through their propagation from the Sun to spacecraft 2. Similar to fast CMEs, most of this variation takes place within Mercury's orbit (14{\%}). Slower CMEs tend to undergo the least amount of variation in speeds during their propagation in the inner heliosphere as the average decrease in speeds for them is only 5{\%}. For fast CMEs, the speed decreases by a factor of 3.8 per AU during their propagation from the Sun to spacecraft 2 while this factor is 3.1 per AU and 2.8 per AU for intermediate and slow CMEs respectively. We find that for CMEs with initial speeds \textgreater 700 km\,s$^{-1}$, there is still significant deceleration past Mercury's orbit.

\begin{figure*}[htbp!]
  \centering
        \includegraphics[width=1.0\linewidth]{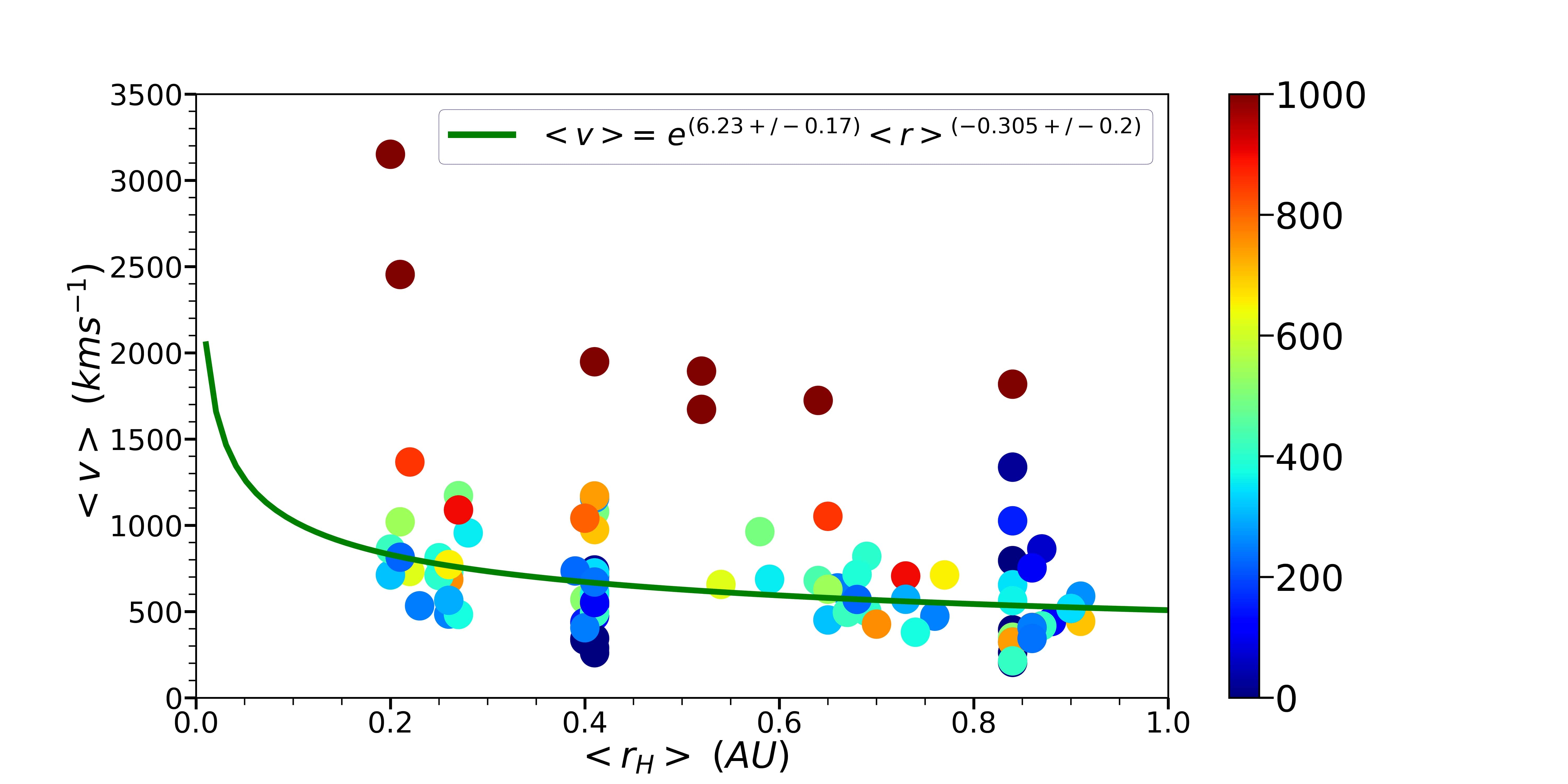}
        \caption{Average transit speed (in km\,s$^{-1}$) versus average heliocentric distance (in AU) plotted along with the best fit power law (green curve) to the data. The relative difference between the initial CME speed and the solar wind speed measured near 1 AU for that specific conjunction event (in km\,s$^{-1}$) is presented as a second scale in color bar.}
         \label{fig:transit}
  \end{figure*}

We also examine the radial evolution of the average transit speed of CMEs. We use the heliocentric distances of the two observation points and the time interval for the transit of the shock/discontinuity (or ejecta if shock/discontinuity arrival time is not available at one of the spacecraft observing the conjunction) to determine the average transit speed. The average transit speeds are then assigned to the average heliocentric distances (mid-points between the two observation points used to determine the average transit speeds). Using a multilinear robust regression fitting technique in logarithmic space, the best fit power law curve to the average transit speed with average heliocentric distance (in Figure~\ref{fig:transit}) is given by: $<v>$= $508^{+94}_{-79}$ $<r>^{-0.305\pm0.2}$ where $<v>$ is the average transit speed (in unit of km\,s$^{-1}$) and $<r>$ is the average heliocentric distance (in unit of AU). The uncertainties represent the 95{\%} confidence interval associated with the fits. The large uncertainty associated with the radial dependency maybe a direct consequence of sampling of average transit speeds concerning different portions of the CME. From MESSENGER and ACE observations, using the same fitting technique, \citeA{Winslow:2015} found a much faster fall-off of the maximum shock speed with heliocentric distance: $<v_{max}>$= $464.05^{+23.79}_{-22.63}$ $<r>^{-0.45\pm0.09}$, although their results were not for CMEs measured in conjunction but obtained statistically for all CMEs measured at Mercury and ACE during 2011-2014. In addition, they used the transit speeds between the Sun and Mercury as the estimated speed at Mercury rather than the speed at the mid-point, as done here. 

\justify

Analyzing the plot, it is evident that events with the highest differences between the initial CME speeds and solar wind speeds measured near 1 AU have a steeper fall-off of average transit speed with distance compared to the rest. Our observations also suggest that deceleration does not become negligible past Mercury's orbit, though the rate of deceleration becomes considerably smaller. 

\subsection{CME Deceleration/Acceleration} \label{ssec:profile2}

\justify

For testing real-time space weather prediction models \cite <e.g.,>[and others]{Siscoe:2006}, it is important to have the exact arrival times of a CME at multiple locations in the inner heliosphere. However, the deceleration/acceleration process is one of the key challenges in predicting the timing of CME arrival at Earth. With scarcity of plasma measurements at sub-1 AU heliocentric distances, it has not yet been possible to construct an average CME deceleration/acceleration profile. That encouraged us to use the estimated impact speeds at Mercury and Venus from the DBM in an attempt to present an average deceleration/acceleration profile of CMEs in the inner heliosphere. CMEs with propagation speeds above that of the ambient solar wind tend to decelerate, while slow ones with propagation speeds below the solar wind speed, get accelerated up to the solar wind speed. This general trend is clearly visible by comparing the fast and slow CMEs in Figure~\ref{fig:fast} and Figure~\ref{fig:slow} respectively, although there is significant event-to-event variability. Previous studies have indicated the deceleration to stop at different heliocentric distances: within Mercury's orbit \cite{Liu:2013} or at 0.76 AU \cite{Gopalswamy:2001b} or anywhere between 0.3 AU and 1 AU \cite{Reiner:2007,Winslow:2015}.

\justify   

\begin{figure*}[htbp!]
  \centering
        \includegraphics[width=1.0\linewidth]{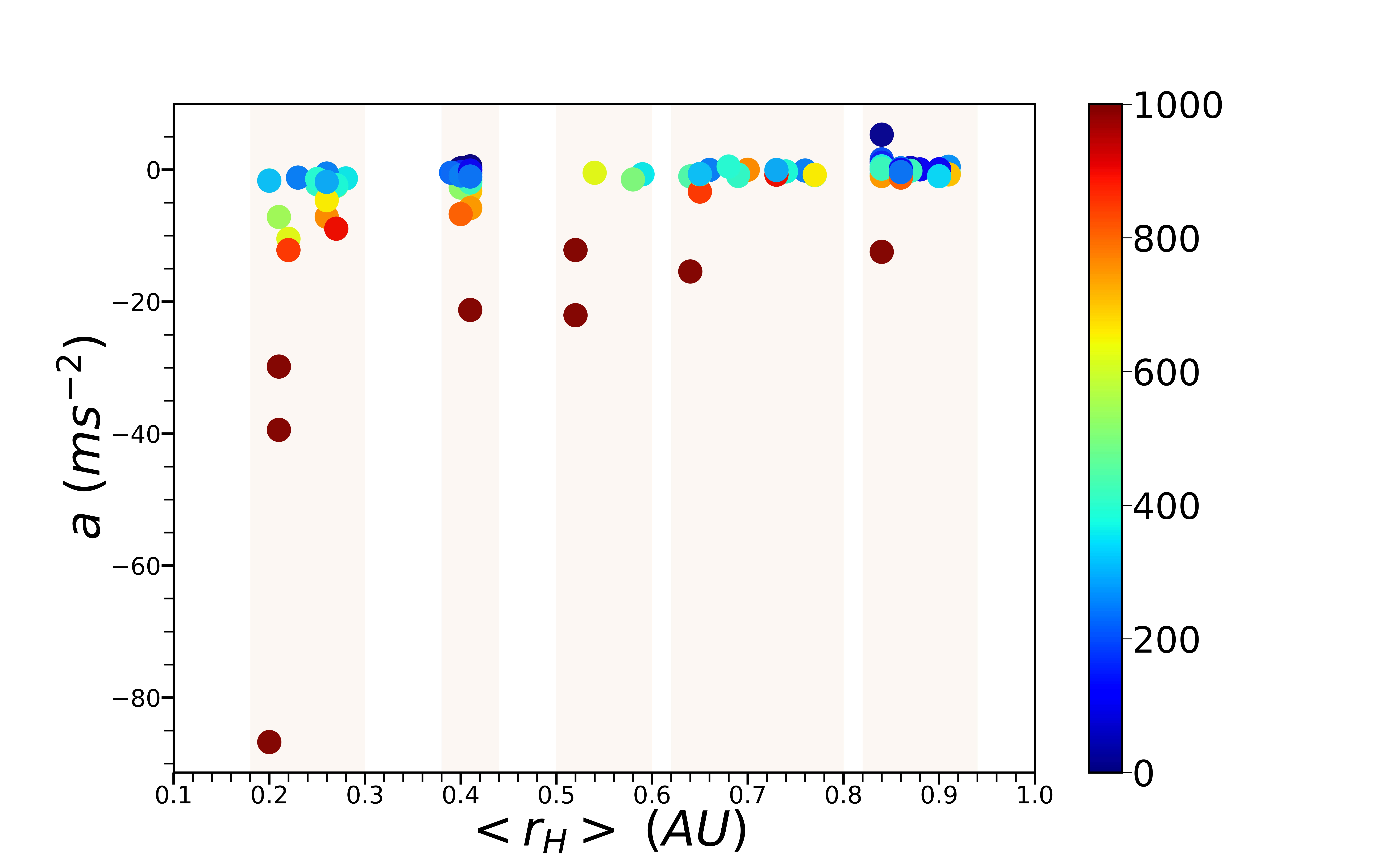}
        \caption{Distribution of average CME deceleration/acceleration (in m\,s$^{-2}$) versus average heliocentric distance between the two observation points (in AU) with the relative difference between the initial CME speed and the solar wind speed measured near 1 AU for that specific conjunction event (in km\,s$^{-1}$) as a second scale in color bar. The heliocentric distance here represents the average distance of the points between which the average CME deceleration/acceleration is estimated. Shaded regions (from left to right): 0.18-0.30 AU [Sun-MESSENGER], 0.38-0.44 AU [Sun-\textit{Venus Express}], 0.50-0.60 AU [MESSENGER-\textit{Venus Express}], 0.62-0.80 AU [MESSENGER-STEREO/\textit{Wind}/ACE], 0.82-0.94 AU [\textit{Venus Express}-STEREO/\textit{Wind}/ACE].}
         \label{fig:A}
  \end{figure*}

Figure~\ref{fig:A} shows the distribution of average CME deceleration/acceleration (calculated as $\frac{V_{2} - V_{1}}{\Delta t}$) against the average heliocentric distance between the two points where we have speed observations/measurements/estimations, with the relative difference between the initial CME speed and the solar wind speed measured near 1AU for that specific conjunction event as a second variable. We use the initial CME speed, average estimated impact speed at Mercury and Venus from the DBM, and the maximum CME speed measured at STEREO/\textit{Wind}/ACE to determine the average deceleration/acceleration of CMEs. The scatter plot highlights that the rate of deceleration/acceleration is predominantly driven by the relative speed difference between the CME and the solar wind, resembling assumptions from previous studies \cite <e.g.>[]{Gopalswamy:2000,Vrsnak:2004,Yashiro:2004} that beyond 20 R$_{s}$, CME dynamics is solely governed by the drag force, hence depends on the quantity (V$_{CME}$ - V$_{SW}$)$^{2}$ \cite <e.g.,>[]{Cargill:2004}. Figure~\ref{fig:A} uses the average heliocentric distance as the mid-point between the two points used to calculate the deceleration/acceleration. For example, a heliocentric distance of 0.27 AU in Figure~\ref{fig:A} represents the average CME deceleration/acceleration between the first point of CME observation (at the Sun: 0.09 AU) and the second point of \textit{in situ} measurement (at MESSENGER: 0.44 AU). We observe the CMEs to undergo the maximum amount of deceleration in their propagation from the Sun to MESSENGER with an average (median) deceleration of 10.9 (2.5) m\,s$^{-2}$. With increasing heliocentric distance, this deceleration decreases but does not become negligible, until beyond Venus's orbit, with an average (median) deceleration of 7.4 (1.5) m\,s$^{-2}$ from MESSENGER to \textit{Venus Express}, and 0.2 (0.1) m\,s$^{-2}$ from \textit{Venus Express} to STEREO/L1. This provides additional evidence for the argument that CME deceleration continues past the orbit of Mercury \cite <e.g.,>[]{Winslow:2015}, at least to Venus's orbit. In the case of some slow CMEs [with initial (measured CME speeds near 1 AU) CME speeds of 177 (340), 282 (375), 313 (350), 318 (460), and 643 (780) km\,s$^{-1}$], however, we do observe an actual increase in speeds during their propagation from the Sun to near 1 AU. A similar trend was observed for only one intermediate CME [with initial (measured CME speed near 1 AU) CME speed of 716 (720) km\,s$^{-1}$] and the relative speed difference between the Sun and near 1 AU was negligible and well within the uncertainty of the various measurements.  

\subsection{Magnetic Field Intensity} \label{ssec:profile3}

\justify

We list the mean and the standard deviation of the maximum magnetic field intensity within each of the substructures (sheath and ejecta) measured at MESSENGER, \textit{Venus Express}, and STEREO/\textit{Wind}/ACE in Table~\ref{tab:B}. As expected, the values of the magnetic field intensities decrease from MESSENGER to near 1 AU. We also list the ratios between the maximum magnetic field intensity measured in the sheath to that of the ejecta in Table~\ref{tab:B}. We find this ratios to remain relatively constant even with increasing heliocentric distance: 1.04 at MESSENGER, 1.04 at \textit{Venus Express}, and 1.18 at STEREO/\textit{Wind}/ACE, consistent with the statistical findings obtained for non-conjunction events in \citeA{Janvier:2019}. It is important to note here that these ratios should only be considered rough estimates, as CME distortion and the possibility of the observing spacecraft traversing through different structures within the CME were not considered in the analysis. However, interpreting multi-spacecraft observations of the same CME can be really difficult, as seen from the analysis of six possible multi-spacecraft observations by STEREO-A, \textit{Wind}, and STEREO-B, listed by \citeA{Kilpua:2011}. For the CME in the interval of 21-22 May 2007, both STEREO-A and STEREO-B observed this CME at a longitudinal separation of 9$^{\circ}$. Even for this small amount of separation, clear differences were reported in the magnetic field structures from the STEREO observations. At STEREO-B, the flux rope strucutre was clearly evident, while at STEREO-A, the CME strucutre was more complex. The maximum magnetic field measured at the spacecraft differed by a factor of $\sim$2 (9.9 nT at STEREO-A and 17.6 at STEREO-B). They presumed that the magnetic cloud was crossed through the center by STEREO-B while STEREO-A encountered the flank. However, for the 19 November 2007 CME, even for a large longitudinal separation of 40.8$^{\circ}$, the maximum magnetic field measured at the spacecraft were more comparable (12.3 nT at STEREO-A and 17.2 nT at STEREO-B). \citeA{Farrugia:2011} conducted a comprehensive analysis of this event and found almost similar fitted magnetic field strength at STEREO-B and \textit{Wind} (21.9 nT at STEREO-B and 23.2 nT at \textit{Wind}). \citeA{Lugaz:2018} analyzed 35 CMEs with \textit{Wind} and ACE observations and highlighted that differences in the total magnetic field strength from one observing spacecraft to another at 1 AU varies slowly with increasing non-radial separation compared to the magnetic field components. They indicated that even with non-radial separation of 4-7$^{\circ}$ between the measuring spacecraft, almost zero correlation can be expected in the observed magnetic field components, where as in the case of the total magnetic field strength, this separation range is clearly higher (14-20$^{\circ}$).

\begin{table}[htbp]
  \centering
  \caption{Mean and the Standard Deviation of the Maximum Magnetic Field Intensities (B$_{max}$) within the Sheath and the Magnetic Ejecta (ME) at MESSENGER, \textit{Venus Express}, and STEREO/\textit{Wind}/ACE. The Standard Deviations Represent 1-$\sigma$ Uncertainty.}
    \begin{tabular}{cccc}
    B$_{max}$           & MESSENGER       & \textit{Venus Express} & Near 1 AU \\
    Sheath (in nT)          & 99$\pm$63             & 28$\pm$11              & 18$\pm$8 \\
    Sheath:Fast CMEs (in nT)          & 118$\pm$84             & 36$\pm$12              & 24$\pm$8 \\
    Sheath:Slow CMEs (in nT)          & 69$\pm$26             & 20$\pm$5              & 14$\pm$5 \\
    ME (in nT)          & 95$\pm$47             & 28$\pm$18              & 16$\pm$7  \\
    ME:Fast CMEs (in nT)          & 109$\pm$64             & 47$\pm$23              & 20$\pm$10  \\
    ME:Slow CMEs (in nT)          & 75$\pm$24             & 20$\pm$7              & 13$\pm$4  \\
    B$_{max,sheath}$/B$_{max,ejecta}$   & 1.04$\pm$0.27            & 1.04$\pm$0.41            & 1.18$\pm$0.45 \\
    \end{tabular}%
  \label{tab:B}%
\end{table}%

\begin{figure*}[htbp!]
  \centering
        \includegraphics[width=1\linewidth]{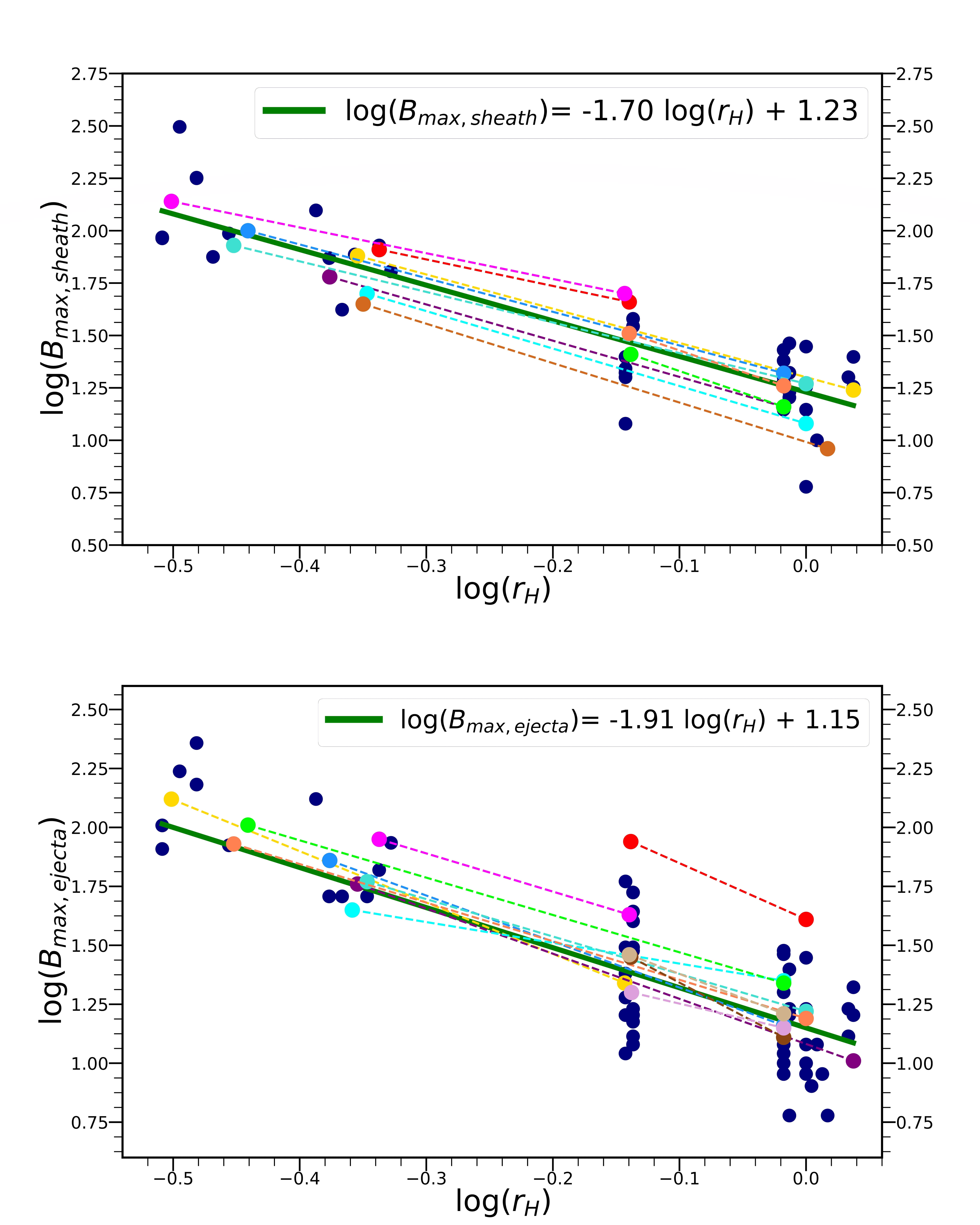}
        \caption{Maximum magnetic field strength (in nT) measured at MESSENGER, \textit{Venus Express}, STEREO, and \textit{Wind}/ACE plotted as a function of heliocentric distance (in AU). The green line represents the best linear fit to this data set. (top) Maximum magnetic field strength in the sheath, (bottom) maximum magnetic field strength in the ejecta. The color-coded lines connect measurements of the same CME at two spacecraft of those events for which the separation was less than 10$^{\circ}$ (10 events, top panel) or 6$^{\circ}$ (12 events, bottom panel).}
         \label{fig:SE}
  \end{figure*} 

\justify  

We plot the maximum magnetic field strength measured in the sheath and the ejecta (Figure~\ref{fig:SE}) as a function of heliocentric distance using MESSENGER, \textit{Venus Express}, STEREO, and \textit{Wind}/ACE measurements. We use a multilinear robust regression fitting technique in logarithmic space to fit a power law curve to our data set. Robust regression uses iteratively reweighted least squares with a bisquare weighting function to ensure that less weight is attributed to outliers than in ordinary least squares fitting. The best fit power law to the maximum magnetic field strength data is: B$_{max,sheath}$= $17.12^{+2.37}_{-2.09}$ $r^{-1.7\pm0.21}$ , B$_{max,ejecta}$= $14.01^{+1.95}_{-1.71}$ $r^{-1.91\pm0.25}$ where B$_{max,sheath}$ and B$_{max,ejecta}$ are the maximum magnetic field strength in nT measured in the sheath and the ejecta respectively and r is in unit of AU. The uncertainties represent the 95{\%} confidence interval associated with the fits.

\justify   

The maximum magnetic field strength in the sheath falls off as $\sim r^{\alpha}$ [see \citeA{Dumbovic:2018} for a discussion of logarithmic decrease of magnetic field and increase in size of CMEs and their relation] where $\alpha$ is -1.7$\pm$0.21. In the case of the ejecta, the maximum magnetic field strength falls off as $\sim r^{\alpha}$ where $\alpha$ is -1.91$\pm$0.25, which is in reasonable agreement with previous theoretical considerations \cite{Demoulin:2009} and empirical fits \cite{Farrugia:2005,Leitner:2007,Wang:2005,Winslow:2015}. 

\justify  

We can also calculate this drop-off not only in a statistical way using all data, as done above, but using the fact that we have conjunction between two spacecraft for each event. Rather than fitting B$_{max}$ vs r$_{H}$, we calculate the exponential decrease of B$_{max,sheath}$ and B$_{max,ejecta}$ between the two spacecraft for each conjunction event. Using this, we find an average (median) $\alpha$ for the sheath to be -1.55 (-1.60). The 25th-75th percentile range is between -1.85 and -1.29. In the case of the ejecta, average (median) $\alpha$ is -1.75 (-1.65) and 50{\%} of the $\alpha$ values lie in the range between -2.29 and -1.35. Here, we find that the average behavior (obtained as the fit to the data) is not the same as the typical individual behavior. This highlights the fact that there might not be one unique behavior for all CMEs, which is masked by performing fits. It also shows that there can be CME-to-CME variability much larger than the typical reported error bar on these fits as the {95\%} confidence interval for the exponential decrease of the maximum magnetic field strength in the ejecta according to our fit is -2.16 to -1.66. Such large spread in radial dependencies can be a direct consequence of the longitudinal separations between the measuring spacecraft. If longitudinal separation is indeed the primary influencing factor, we expect larger deviations from the regression line for events with higher separations. In Figure~\ref{fig:SE}, we make an attempt to qualitatively represent this correlation between longitudinal separation and event-to-event variability. For small longitudinal separations (less than 10$^{\circ}$), we observe the scaling of the maximum magnetic field strength in the sheath to more or less follow the power law fit equation (see top panel of Figure~\ref{fig:SE}). However, in the case of the maximum magnetic field strength in the ejecta, the trend is not straightforward. For even smaller longitudinal separations (less than 6$^{\circ}$), we find the scaling of the maximum magnetic field strength in the ejecta of a higher number of events to show significant deviation from the regression line (see bottom panel of Figure~\ref{fig:SE}).

\justify

\begin{figure*}[htbp!]
  \centering
        \includegraphics[width=1.0\linewidth]{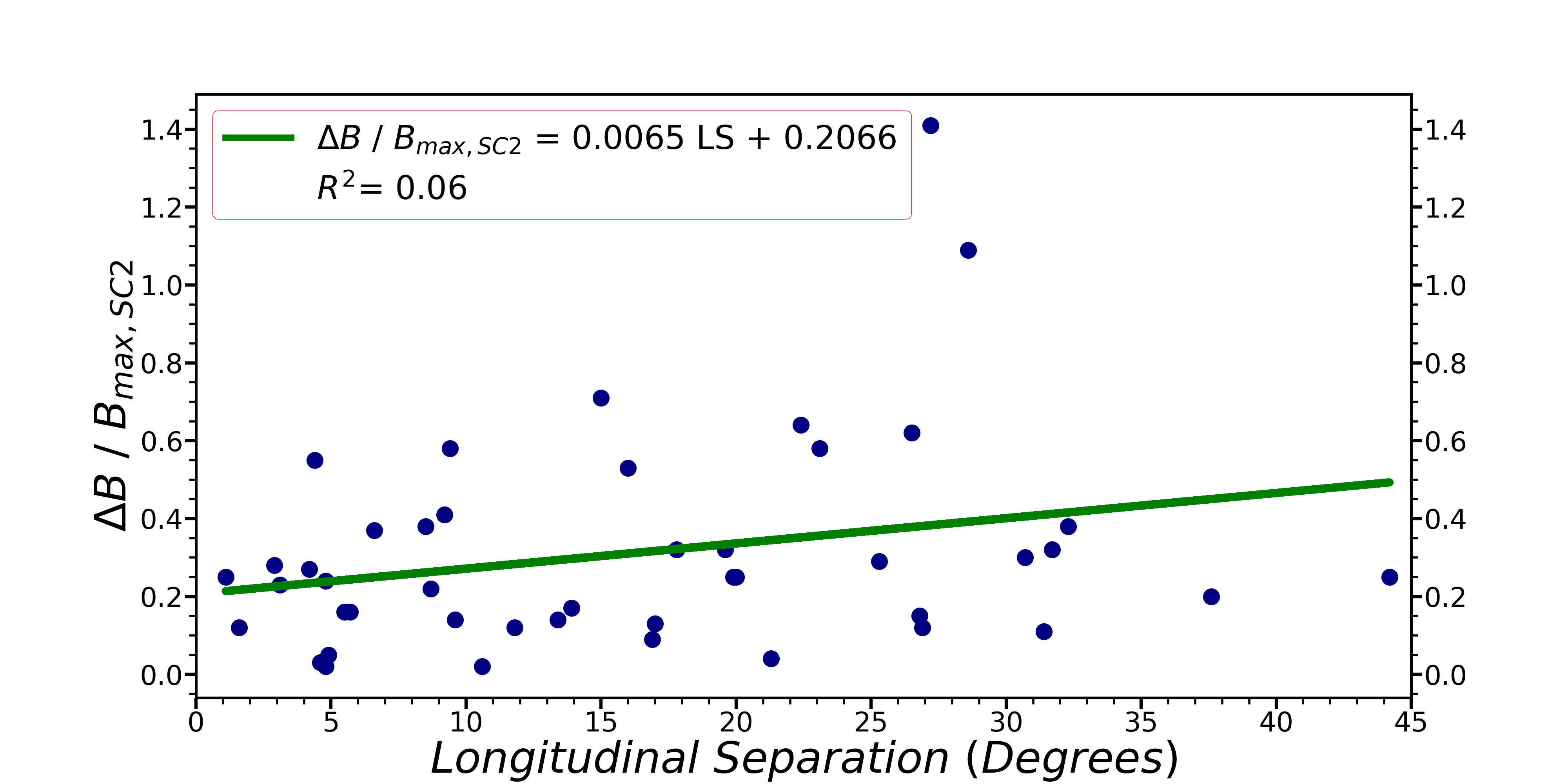}
        \caption{Normalized values of $\Delta$B (for the ejecta, details in the text) plotted as a function of longitudinal separation (in degree). The green line represents the best linear least squares fit to this data set.}
         \label{fig:Err}
  \end{figure*}

This finding prompted us to further explore the possible correlation between longitudinal separations and deviations from the regression line. We make an attempt to quantify it in a statistical manner. Using the maximum magnetic field strength measured in the ejecta at spacecraft 1, we use the scaling constant derived from our robust regression fitting ($\alpha$= -1.91) to find the expected maximum magnetic field strength in the ejecta at spacecraft 2 using the following equation: B$_{max2,expected}$= B$_{max1,measured}$ $R^{-1.91}$, here R is the ratio of positioning of spacecraft 2 (in AU) to positioning of spacecraft 1 (also in AU). Then, we determine the difference between this expected value and the actual measured maximum magnetic field strength in the ejecta at spacecraft 2 and call it $\Delta$B. We normalize the values of $\Delta$B to the measured maximum magnetic field strength in the ejecta at spacecraft 2. We fit the normalized values against longitudinal separations (LS) between the measuring spacecraft, using a linear least squares fitting technique in Figure~\ref{fig:Err}. The fit results confirms the randomness of the maximum magnetic field strength in the ejecta (seen in bottom panel of Figure~\ref{fig:SE}) and finds almost zero correlation with longitudinal separations.

\justify

\begin{figure*}[htbp!]
  \centering
        \includegraphics[width=1.0\linewidth]{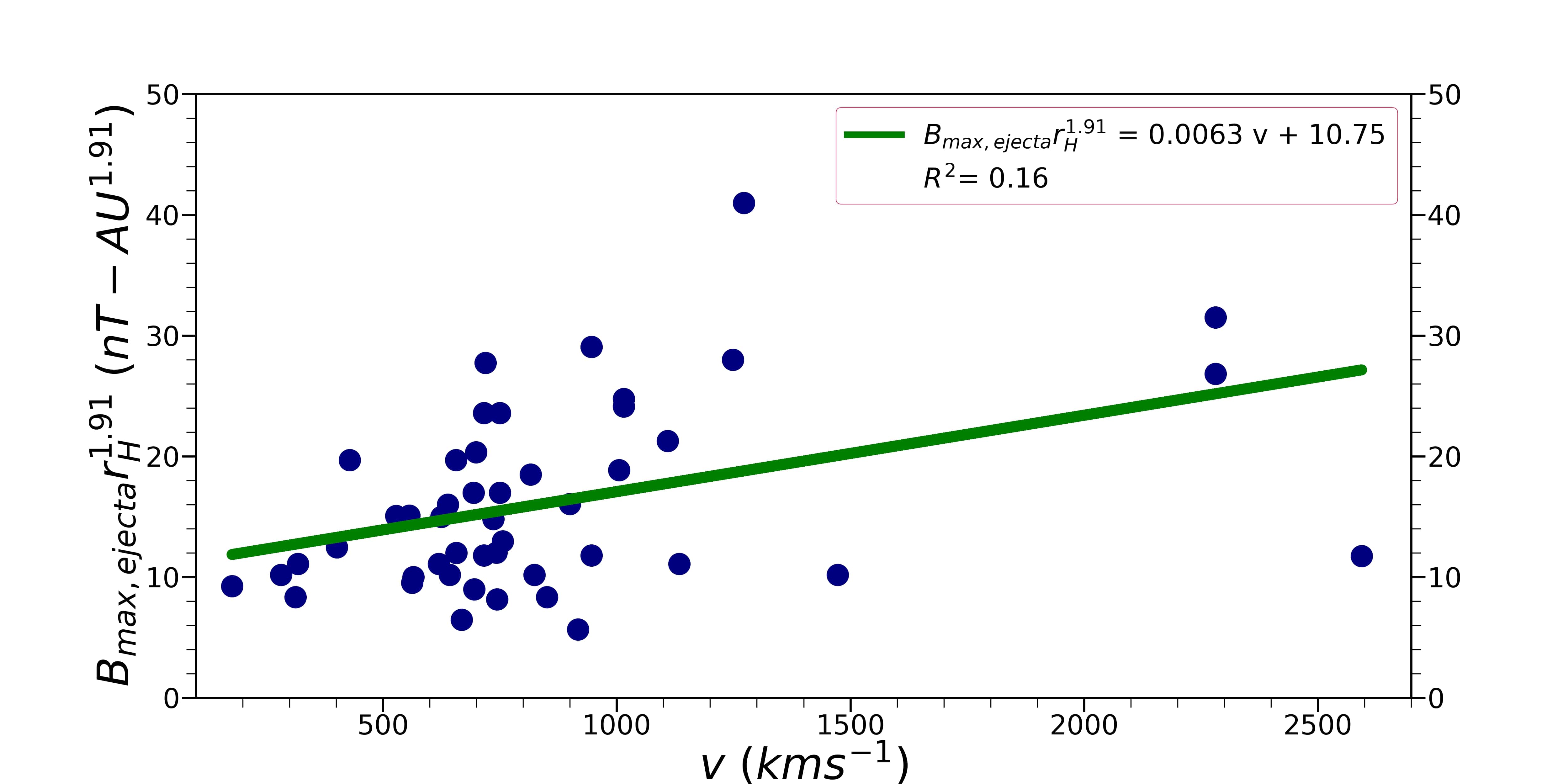}
        \caption{B$_{max,ejecta}$ r$^{1.91}$ (in nT - AU$^{1.91}$) plotted as a function of initial CME speed (in km\,s$^{-1}$). The green line represents the best linear least squares fit to this data set.}
         \label{fig:BsqR}
  \end{figure*}

Next, we attempt to find if the maximum magnetic field strength inside the CME depends on other parameters in addition to the radial distance. However, our B$_{max,ejecta}$ values are measured at different heliocentric distances, not just at 1 AU. Therefore, we use one representative measurement for each conjunction event which is the closest to 1AU and multiply with r$^{\alpha}$ to scale it to 1 AU. Now, all but 5 of our magnetic field measurements are in the range of 0.96-1.09 AU, which we scale back to 1AU. The other 5 measurements are in the range of 0.72-0.73 AU. In this way the uncertainty associated with the scaling parameter ($\alpha$, found from the multilinear robust regression) shall have the least effect on the analysis. Based on past work \cite{Demoulin:2009}, we expect this quantity B$_{max,ejecta}$ r$^{\alpha}$ to be a constant of CME propagation. We examine the dependence of this quantity on the initial CME speed in Figure~\ref{fig:BsqR}. We perform a linear least squares fitting of B$_{max,ejecta}$ r$^{1.91}$ against the initial CME speed. The correlation is very weak and the best fit linear regression line has a minimal slope which implies that this quantity is fairly independent of the initial CME speed. The slope is 3 times less (0.0063 compared to 0.0189) than that found by \citeA{Moestl:2014} using measurements only at 1 AU and the average propagation speed from HI observations, as well as the correlation found here is much weaker. Part of the reason for this discrepancy could be that we use the initial (not propagation) speed or because we use scaled magnetic field measurements at various heliocentric distances.   

\subsection{Sheath Duration} \label{ssec:profile4} 

\justify  

\begin{figure*}[htbp!]
  \centering
        \includegraphics[width=1.0\linewidth]{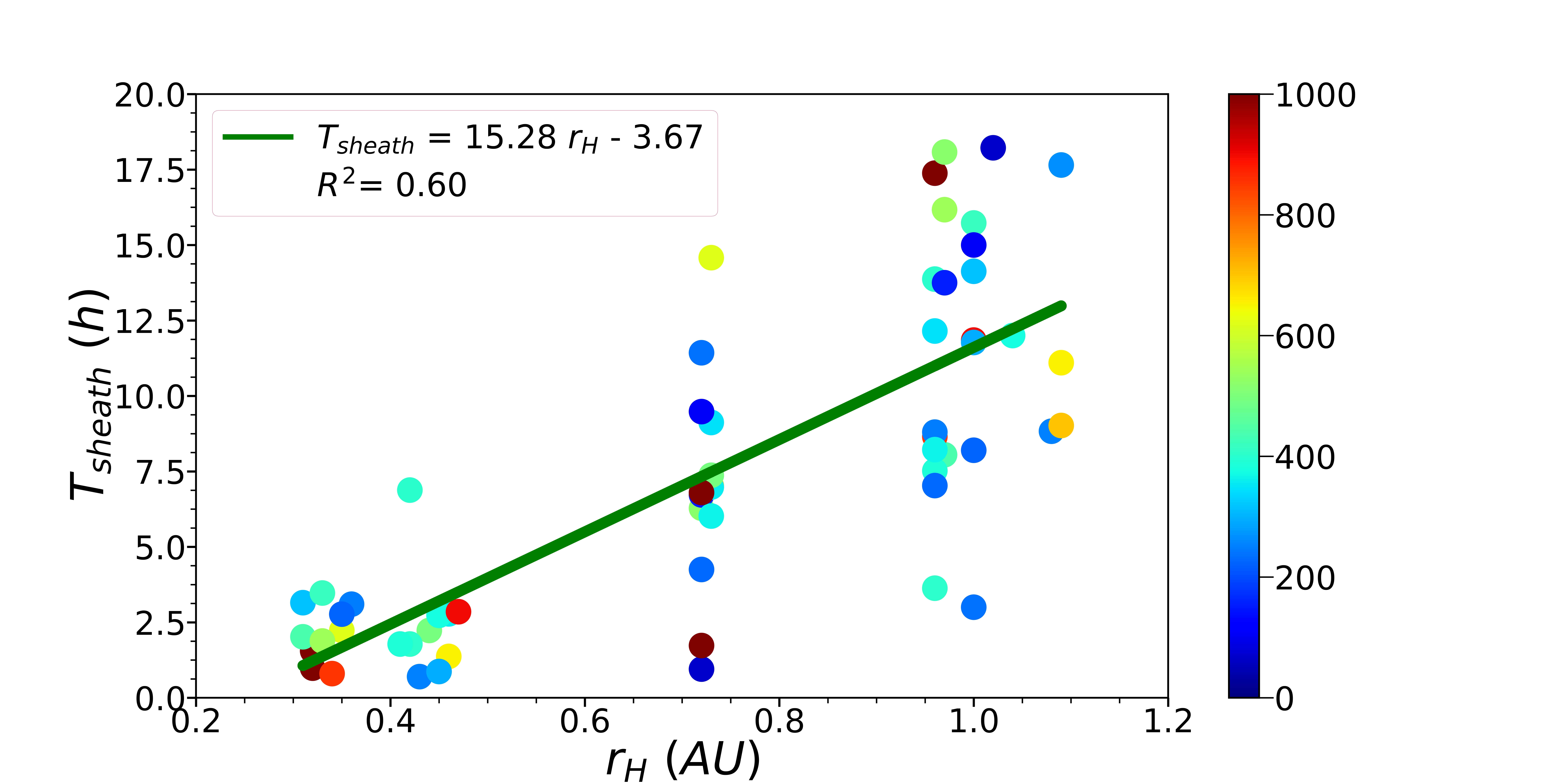}
        \caption{CME sheath duration (in h) plotted as a function of heliocentric distance (in AU) from MESSENGER, \textit{Venus Express}, STEREO, and \textit{Wind}/ACE measurements with the relative difference between the initial CME speed and the solar wind speed measured near 1 AU for that specific conjunction event (in km\,s$^{-1}$) as a second scale in color bar. The coefficient of determination and the best linear least squares fit (green straight line) to the data are shown.}
         \label{fig:Sheath}
  \end{figure*}

We use the CME shock/discontinuity and the ejecta arrival times to estimate the sheath duration at MESSENGER, \textit{Venus Express}, STEREO, and \textit{Wind}/ACE. We use a linear least squares fitting of the sheath duration (in hours) against heliocentric distance (in AU) and find the relationship between them to be T$_{sheath}$= 15.28 r$_{H}$ - 3.67 (Figure~\ref{fig:Sheath}). R$^{2}$ of the linear regression line is 0.60 which represents a decent correlation between the parameters. The large positive slope indicates a linear increase in sheath duration with increasing heliocentric distance. This confirms commonly held knowledge that the CME sheath increases with heliocentric distance. We do not have enough data to test whether this increase is anything but linear. It is important to note that the large intercept of -3.67 of the fit implies that the formation of the sheath starts around 0.24 AU. However, we find the sheath duration on average to be fairly independent of the initial CME speed throughout the inner heliosphere. At Mercury, we find on average the sheath duration of fast CMEs to be 1.74 hours and slow CMEs to be 2.22 hours. At Venus, the average sheath duration for fast and slow CMEs increase by a factor of $\sim$4 and $\sim$3 respectively compared to Mercury with an avearge sheath duration of 7.30 and 6.57 hours for fast and slow CMEs respectively. Near 1 AU, the average sheath duration for fast and slow CMEs are 13.74 hours and 10.98 hours respectively. This finding runs contrary to commonly held expectation that slower CMEs have larger sheaths than fast CMEs, though our sample size for the analysis is small. This belief is mostly based on comparison with magnetosheaths. However, as explained by \citeA{Siscoe:2008}, CME sheaths differ in many respects from magnetosheaths. It should also be noted that if sheath thickness is considered rather than sheath duration, according to this finding, fast CMEs have even thicker sheaths. It is an interesting revelation contradicting the results of \citeA{Russell:2002,Savani:2011b} where stand-off distance of the shock (the separation distance from the obstacle boundary to the shock) decreases with Mach number (ratio of the speed of the shock in the solar wind frame to the fast-magnetosonic speed) meaning slower CMEs, rather than fast ones should have thicker sheaths. Important thing to note that we do not take into consideration (for lack of measurements) the CME shape, radius of curvature, and the distance to the CME nose for this analysis. This finding does not take advantage of the conjunction (same analysis could be done even if these events were not in conjunction). For pairs of measurements done in conjunction, we observe the sheath duration to increase by a factor of 14.90, 6, and 13.90 per AU for fast, intermediate, and slow CMEs respectively. However, for the fast and slow CMEs especially, these averages are clearly driven by some irregular values arising from the effect of longitudinal separation between the observing spacecraft. 

\section{Case Study of the 3 Nov 2011 Event} \label{sec:case study}

\subsection{CME Propagation} \label{ssec:propagation}

\justify

\citeA{Good:2015,Good:2018} studied this same CME event to examine the change in the expansion speed, self-similarity nature and flux rope fit. However, our primary focus is on the relation between the expansion speed and change of B$_{max}$ as well as the development of the sheath. We also present it as an example to highlight the processes we conducted to build the conjunction catalog.

\justify

\begin{figure*}[htbp!]
  \centering
        \includegraphics[width=1.0\linewidth]{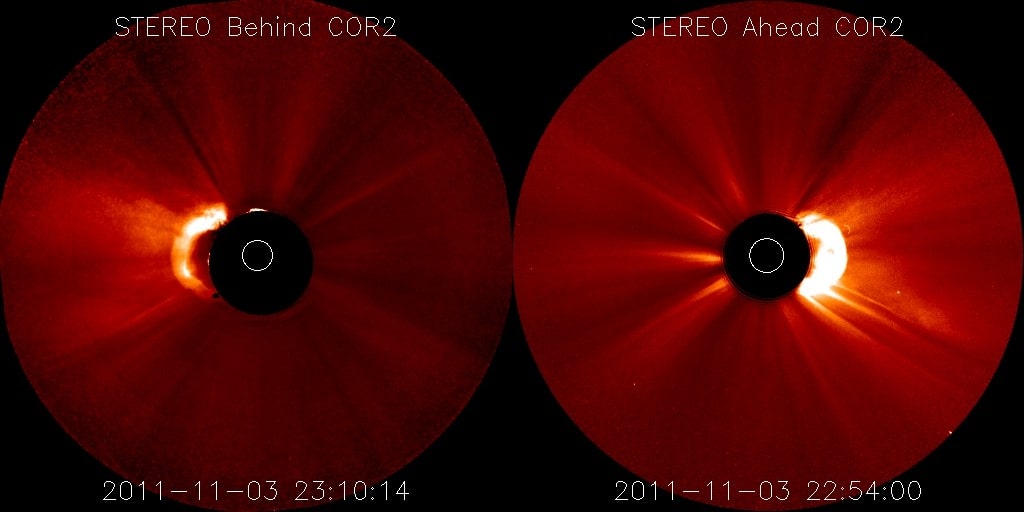}
        \caption{COR2 (outer coronagraph) observations of the 3 November 2011 CME event from STEREO-B (left) and STEREO-A (right).}
         \label{fig:COR2}
  \end{figure*}

The CME counterpart first appeared in the LASCO C2 field of view as a halo on 3 November 2011 at 23:48 UT. This CME emerged from the Sun's western limb at 22:54 UT on 3 November when viewed from STEREO-A and from the Sun's eastern limb at 23:10 UT on 3 November when viewed from STEREO-B (Figure~\ref{fig:COR2}). The listed non-linear speed at 20 R$_{s}$ for the CME in the CDAW catalog is 946 km\,s$^{-1}$.

\justify

We expect the non-linear speeds listed in the CDAW catalog to produce comparable estimated arrival times to observed arrival times at different heliocentric distances. Even under the assumption of minimum drag (drag parameter of $0.1\times10^{-7} km^{-1}$), the DBM estimated arrival time of this CME at MESSENGER was not in reasonable agreement with the listed arrival time at MESSENGER by \citeA{Winslow:2015}. We attribute this discrepancy in timing to the listed non-linear speed at 20 R$_{s}$ being an estimate from coronagraph observations. As a result, we use two known constraints: listed CME onset time at 20 R$_{s}$ in the CDAW catalog and the listed shock arrival time at MESSENGER under the assumption of minimum drag to estimate the speed of the CME at 20 R$_{s}$. Under these conditions, the DBM requires the CME speed at 20 R$_{s}$ to be at least 1140 km\,s$^{-1}$. With this minimum speed of 1140 km\,s$^{-1}$ at 20 R$_{s}$, we also estimate the arrival time of the CME at \textit{Venus Express} under the assumption of minimum drag. We find the DBM estimated arrival time at \textit{Venus Express} to be in perfect agreement with the time listed by \citeA{Good:2016} which verifies our assumption about the listed non-linear speed at 20 R$_{s}$ for this CME in the CDAW catalog to be not entirely accurate. This process also eliminates a second coronagraphic CME candidate (LASCO onset time of 01:25 UT on 4 November 2011 with 716 km\,s$^{-1}$ as the listed non-linear speed at 20 R$_{s}$) as it is considerably slower than the chosen event. However, the observed complex \textit{in situ} signatures of the magnetic field at \textit{Venus Express} prompted us to query about possible interaction between these two CME candidates. Interaction of the two candidate CMEs as they propagate is unlikely, especially at MESSENGER and STEREO-B, in light of the clear magnetic ejecta signatures, quite different from the possible \textit{in situ} manifestations of CME-CME interactions \cite <see>[]{Lugaz:2017b}. The coronal speed of the two CMEs also make their interaction highly unlikely.

\justify

\begin{figure*}[htbp!]
  \centering
        \includegraphics[width=1.0\linewidth]{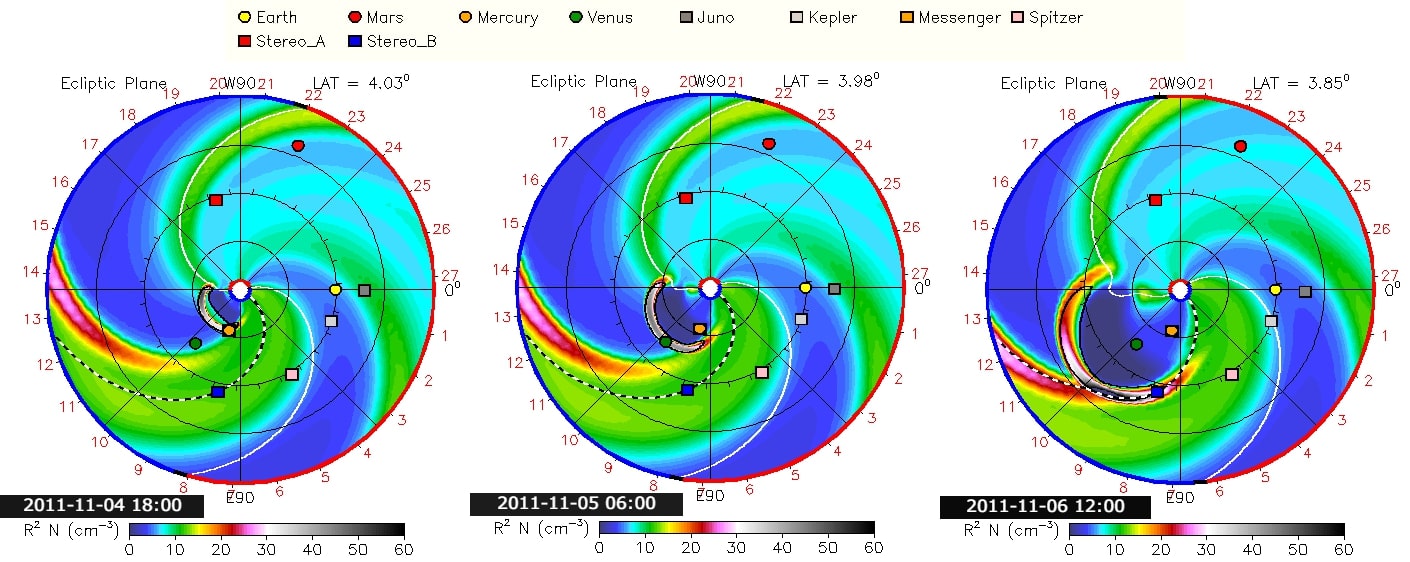}
        \caption{WSA-ENLIL model simulation with CME cone extension of the 3 November 2011 CME: when it reached Mercury (left), Venus (middle) and STEREO-B (right).}
         \label{fig:MSTB}
  \end{figure*}

Figure~\ref{fig:MSTB} shows the positioning of the inner heliosphere planets and spacecraft and WSA-ENLIL model with cone extension \cite{Odstrcil:2003,Odstrcil:2004} simulated propagation of the CME at three different time steps of interest: at Mercury, Venus, and STEREO-B. The simulations are illustrations of the CME arrival at the points of interest and are not meant for precise timing of arrival. These simulations are taken from SWRC CATALOG of Space Weather Database Of Notifications, Knowledge, Information (DONKI). Initial CME speed of 1100 km\,s$^{-1}$ and half angular width of 65$^{\circ}$ was used for this simulation. 

\justify

\begin{figure*}[htbp!]
  \centering
        \includegraphics[width=1.0\linewidth]{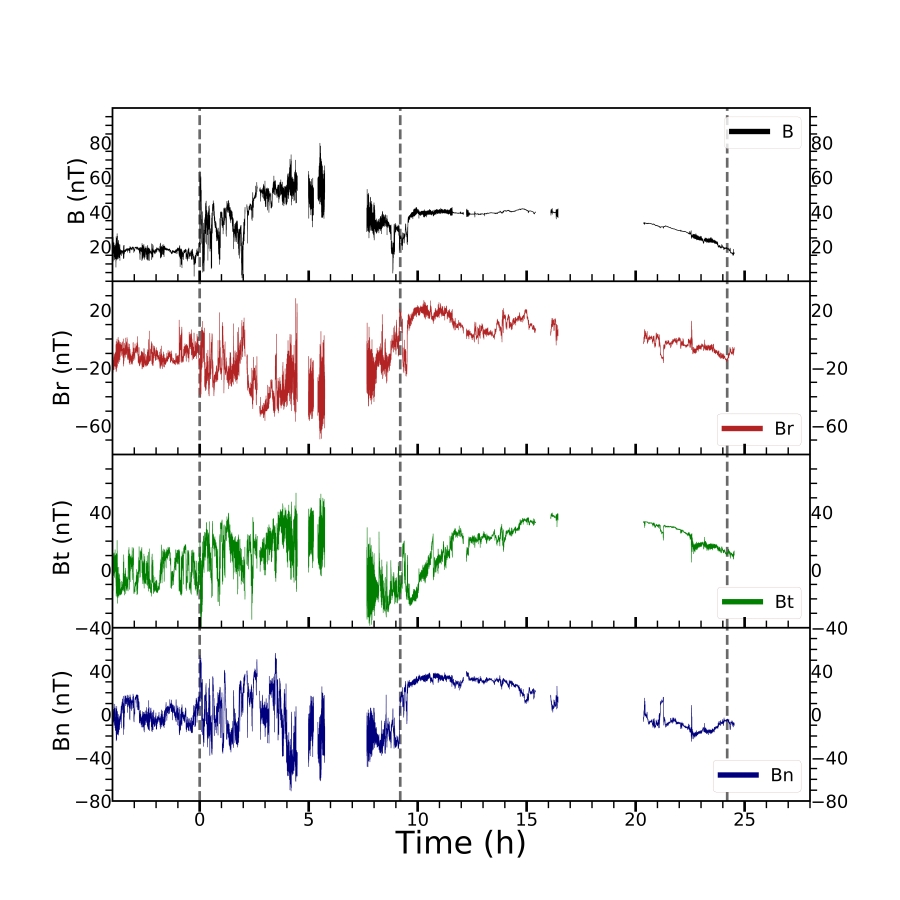}
        \caption{MESSENGER measurements of the CME on 4-5 November 2011. The four panels show magnetic field data in RTN coordinates. Vertical grey dashed lines (from left to right) denote the crossing time of the CME shock (t=0 h), magnetic ejecta, and CME end. The data gap corresponds to MESSENGER's passage through Mercury's magnetosphere.}
         \label{fig:MER}
  \end{figure*}

Due to the close longitudinal alignment of the inner solar system planets at the time, about $\sim$15 hours after the CME eruption, the shock arrived at MESSENGER on 4 November at 15:09 UT (observed in the magnetic field data) with estimated impact speed of 834 km\,s$^{-1}$ (from the DBM). The heliocentric distance of MESSENGER was 0.44 AU. The signatures of the magnetic flux rope are observed at MESSENGER from 00:21 UT to 15:21 UT on 5 November with the arrival of the leading edge, $\sim$9.2 hours after the shock arrival. Using 1 s high-resolution data from the MAG on-board the MESSENGER spacecraft, the magnetic field profile of the sheath and the ejecta are presented in Figure~\ref{fig:MER}. In the plot, t=0 h corresponds to the shock arrival. Intervals when MESSENGER traversed the magnetosphere of Mercury are excluded. The start time and end time of the ejecta at MESSENGER were estimated to lie at the start period of the B$_{n}$ rotation (t=9.2 h) to the period after which the magnetic field returns to its pristine condition (t=24.2 h). Our choice of the ejecta boundaries is consistent with \citeA{Winslow:2015} except the start time of the ejecta, which is significantly different. They define the start of the ejecta at t=2.23 h and end of the ejecta at t=24.5 h. The reason behind adjusting the ejecta boundaries is that \citeA{Winslow:2015} defined the boundaries using only magnetic field measurements. In our case, we have the knowledge that this same event is also observed at STEREO-B where we have plasma measurements. So, we use both magnetic field and plasma measurements at STEREO-B to determine the CME sheath and then back-tracked to MESSENGER.

\justify

\begin{figure*}[htbp!]
  \centering
        \includegraphics[width=1.0\linewidth]{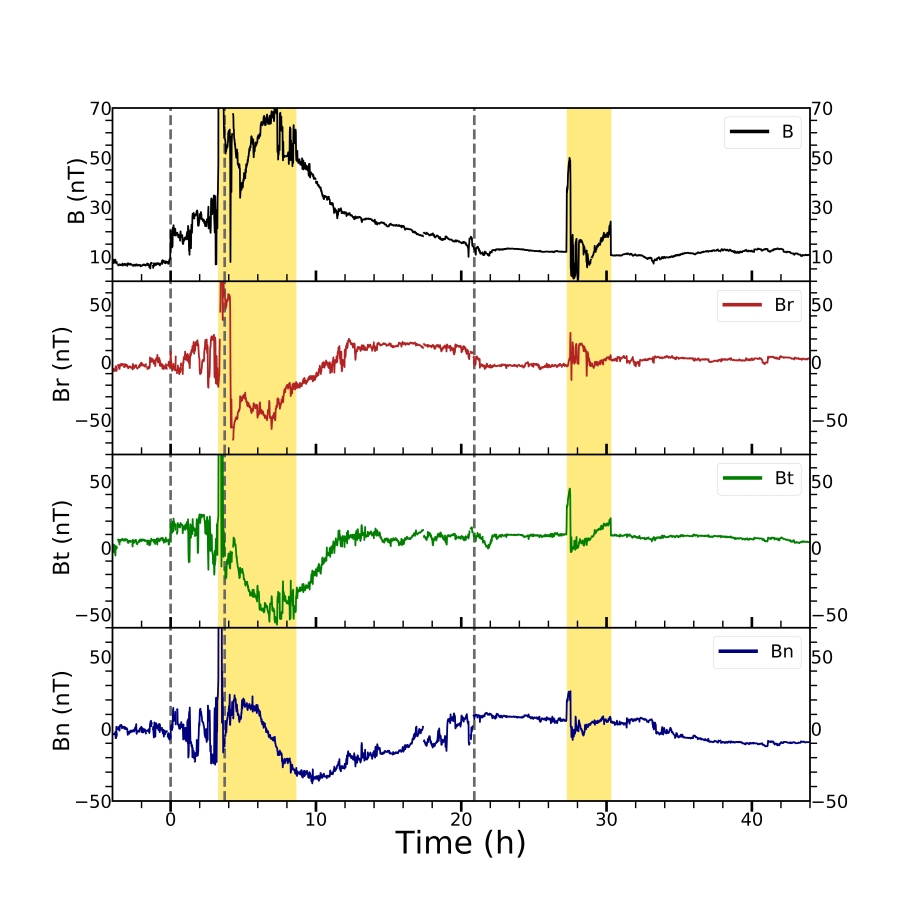}
        \caption{\textit{Venus Express} measurements of the CME on 4-6 November 2011. The four panels show magnetic field data in RTN coordinates. Vertical grey dashed lines (from left to right) denote the crossing time of the CME shock (t=0 h), magnetic ejecta, and CME end. Yellow shaded regions represent the two magnetospheric crossings of \textit{Venus Express} during the event.}
         \label{fig:V}
  \end{figure*}

The shock signature was observed in the MAG on-board the \textit{Venus Express} spacecraft on 5 November 3:42 UT with estimated impact speed of 768 km\,s$^{-1}$ (from the DBM). The heliocentric distance of \textit{Venus Express} was 0.73 AU. We define the ejecta boundaries in the same manner as MESSENGER. The magnetic flux rope is approximated to start at 07:25 UT on 5 November and end at 00:36 UT on 6 November. Figure~\ref{fig:V} shows the magnetic field measurements in the sheath and the ejecta at \textit{Venus Express}. In the plot, t=0 h corresponds to the shock arrival. Intervals of induced magnetic field when \textit{Venus Express} was inside the magnetosphere of Venus are shaded. Our choice of the start of the ejecta is consistent with \citeA{Good:2016}. They approximate the ejecta to start anywhere between 3.23 and 7.37 hours after the shock arrival while we define the ejecta boundary to start at t=3.72 h. However, we define the end of the ejecta to be at t=20.9 h where they list the ejecta to end at t=12.12 h. 

\justify

\begin{figure*}[htbp!]
  \centering
        \includegraphics[width=1.0\linewidth]{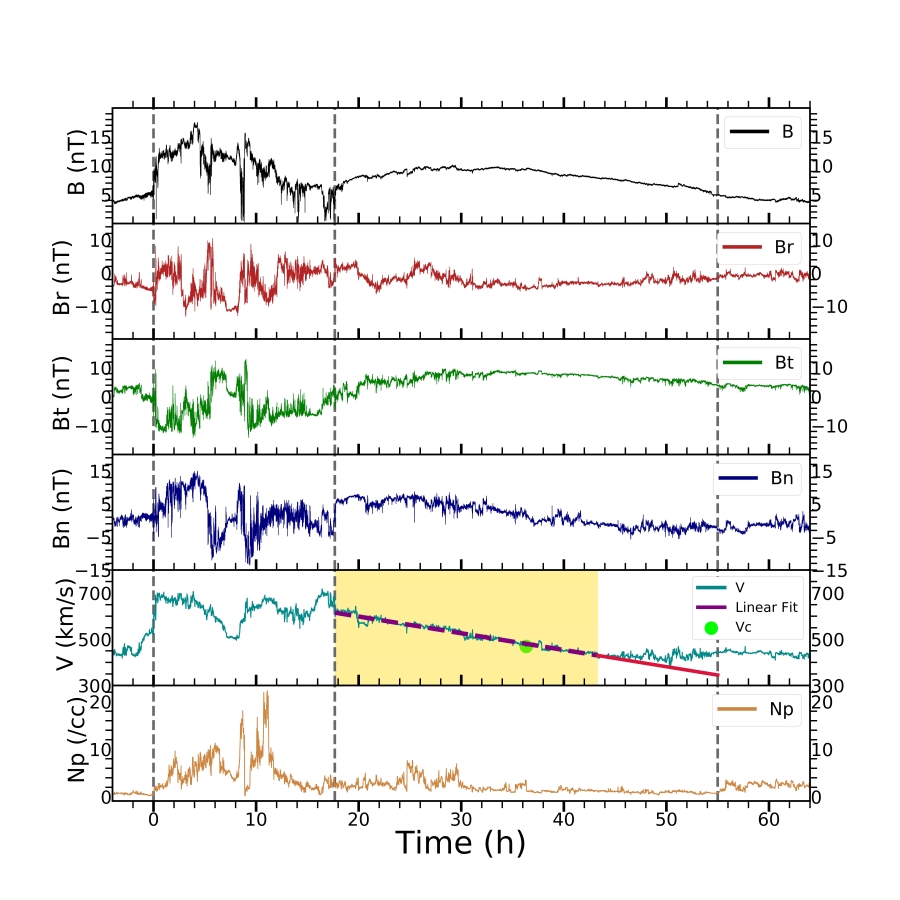}
        \caption{STEREO-B measurements of the CME on 6-8 November 2011. The six panels show magnetic field data in RTN coordinates (first four from top) and plasma measurements (last two). Vertical grey dashed lines (from left to right) denote the crossing time of the CME shock-like discontinuity (t=0 h), magnetic ejecta, and CME end. Yellow shaded region in the fifth panel indicate the time interval where a near linear trend in the solar wind speed is observed in the ejecta.}
         \label{fig:ST}
  \end{figure*}

A weak shock-like discontinuity was observed at 5:11 UT on 6 November at STEREO-B (in the plasma and magnetic field data). The heliocentric distance of STEREO-B was 1.09 AU. At 22:50 UT on 6 November, $\sim$47 hours after the arrival of the flux rope at MESSENGER, the same rope arrived at STEREO-B. The spacecraft encountered the cloud for a period of $\sim$37 hours with the trailing edge arriving on 8 November at 12:11 UT. The leading edge of the flux rope had a speed of 617 km\,s$^{-1}$ while the trailing edge had a speed of 446 km\,s$^{-1}$. The considerable difference between the speeds of the leading edge and the trailing edge represents an expanding magnetic cloud. Using magnetic field data from the IMPACT instrument and plasma data from the PLASTIC instrument on-board STEREO-B, the sheath and the ejecta measurements are presented in Figure~\ref{fig:ST}. In the plot, t=0 h corresponds to the discontinuity arrival. The boundaries of the ejecta at STEREO-B were estimated in a similar way to MESSENGER but in this case we also had \textit{in situ} plasma data. We define the flux rope to start at t=17.67 h and end at t=55h. Our leading edge boundary is consistent with \citeA{Jian:2018}. However, our trailing edge approximation is $\sim$16 h ahead of them.

\subsection{Sheath and Ejecta} \label{ssec:sheath}

In Table~\ref{tab:CB}, we present the total magnetic field intensities measured at the three spacecraft. The ratio of the measured maximum magnetic field strength in the sheath to the measured maximum magnetic field strength in the ejecta at MESSENGER, \textit{Venus Express}, and STEREO-B is found to be 1.86, 0.66, and 1.80 respectively. Except at \textit{Venus Express}, this ratio stayed almost constant, which is consistent with our findings (see Table~\ref{tab:B}). At Mercury, the duration of the sheath is $\sim$ 9 h with MESSENGER observing the ejecta for a period of $\sim$ 15 h. \textit{Venus Express} spent $\sim$ 4 h in the sheath and $\sim$ 17 h in the ejecta. The same sheath is observed by STEREO-B for a $\sim$ 18 h period with the ejecta lasting $\sim$37 h. Looking at the WSA-ENLIL model with cone extension simulated propagation of the CME (see Figure~\ref{fig:MSTB}), it is evident that both MESSENGER and STEREO-B observed the flank of the ejecta while \textit{Venus Express} observed its nose. We believe this to be the reason for a much shorter sheath period than expected at \textit{Venus Express} and a more magnetized ejecta than the sheath.  

\justify

\begin{table}[htbp]
  \centering
  \caption{Total Magnetic Field Intensities within the Sheath and the Ejecta at MESSENGER, \textit{Venus Express}, and STEREO-B. [Note: Avg B$_{sh}$= Average Magnetic Field Strength Measured in the Sheath, Avg B$_{ej}$= Average Magnetic Field Strength Measured in the Ejecta, B$_{m,sh}$= Maximum Magnetic Field Strength Measured in the Sheath,  B$_{m,ej}$= Maximum Magnetic Field Strength Measured in the Ejecta]}
    \begin{tabular}{ccccc}
    Spacecraft      & Avg B$_{sh}$ (nT) & Avg B$_{ej}$ (nT) & B$_{m,sh}$ (nT) & B$_{m,ej}$ (nT) \\
    MESSENGER       & 40              & 35              & 80              & 43 \\
    \textit{Venus Express} &22                 &25                 & 35              &53  \\
    STEREO-B        & 10              & 8               & 18              & 10 \\
    \end{tabular}%
  \label{tab:CB}%
\end{table}%

\subsection{Expansion Speed and Expansion of the Magnetic Ejecta} \label{ssec:MC}

\justify

We want to take advantage of the conjunction to compare the expansion speed and profile measured at 1 AU and the change in the magnetic field strength during propagation. We determine the full expansion speed of the ejecta at STEREO-B through a linear least squares fit of the solar wind speed in the time interval where a near linear trend is present (yellow shaded region in the fifth panel of Figure~\ref{fig:ST}), like \citeA{Gulisano:2010}. Then we use the linear fit to define the speeds at the ejecta boundaries. The expansion speed [$\Delta V = \frac{V_{fit}(t_{LE}) - V_{fit}(t_{TE})}{2}$, LE=leading edge, TE=trailing edge] of the ejecta at STEREO-B is $\sim$ 135 km\,s$^{-1}$.

\justify

Without linear fitting, the expansion speed [$\Delta V = \frac{V_{front} - V_{back}}{2}$] is $\sim$86 km\,s$^{-1}$ for a central speed of 469 km\,s$^{-1}$, an average speed of 493 km\,s$^{-1}$, a front speed of 617 km\,s$^{-1}$ and a back speed of 446 km\,s$^{-1}$. This expansion speed is nearly identical to that obtained by using the relation between leading edge and expansion speed [V$_{exp}$ (km\,s$^{-1}$)= 0.266V$_{LE}$ (km\,s$^{-1}$) - 70.6 (km\,s$^{-1}$), \citeA{Owens:2005}] which is $\sim$ 94 km\,s$^{-1}$. Typical expansion speeds at 1 AU are 40-60 km\,s$^{-1}$ \cite{Richardson:2010} and this CME is found to be expanding fast based on this metrics.
 
\justify

We also estimate the non-dimensional expansion rate $\zeta$ \cite{Gulisano:2010} of the ejecta at STEREO-B (1.09 AU). $\zeta$ is defined as:

$$\zeta = \frac{\Delta V}{\Delta t} \frac{D}{V_c^2},$$ 

\justify

Here $\frac{\Delta V}{\Delta t}$ is the slope of the best linear-fit, D is the distance to the Sun, and V$_{c}$ is the plasma speed measured at the center of the flux rope (see fifth panel of Figure~\ref{fig:ST}). We measure V$_{c}$ to be 469 km\,s$^{-1}$. We find $\zeta$ to be 1.5 which represents a very fast expanding CME as for non-perturbed MCs, typical $\zeta$$\approx$ 0.8 \cite{Gulisano:2010} with the typical spread being $\pm$0.19 \cite{Demoulin:2010}. We examine the variation of this expansion parameter, as \citeA{Demoulin:2009} showed this parameter to stay relatively constant with distance. According to them, the magnetic field in the ejecta fall off as $\sim r^{\alpha}$ \cite <see also>[]{Dumbovic:2018} where $\alpha$$\approx$ -2$\zeta$. Thus for this conjunction event with the value of $\zeta$ being 1.5, we expect the magnetic field in the ejecta to fall off as $\sim r^{-3}$. The maximum magnetic field strength in the ejecta at 0.44 AU (Mercury) is 43 nT, it is 53 nT at 0.73 AU (Venus) and 10 nT at 1.09 AU (STEREO-B). The exponent decrease of the magnetic field is therefore $\alpha_{M,VEx}$ = 0.41 between Mercury and Venus, $\alpha_{VEx,STB}$ = -4.16 between Venus and STEREO-B and $\alpha_{M,STB}$ = -1.61 between Mercury and STEREO-B. We do not consider the $\alpha_{B}$ values for Venus as it encountered the nose of the ejecta whereas Mercury and STEREO-B encountered the flank. The value of $\alpha_{B}$ between Mercury and STEREO-B is smaller than the median found in section~\ref{ssec:profile3}. This shows that the decrease of B is slower than typical inside this CME. We are left with an unexpected result: the CME is found to expand quickly at 1 AU, but the decrease of the magnetic field is consistent with a weaker-than-average expansion. This might be due to the fact that the CME flank is encountered at MESSENGER and STEREO-B. 

\section{Discussion and Conclusions} \label{sec:conclusion}

\justify

In general, the evolution of CME properties in the inner heliosphere are inferred from statistical approaches concerning different CME events measured at different heliocentric distances, because multi-point analysis of CME events are mostly limited to case studies. However, statistical laws are only true on average. Therefore, statistical results have the tendency to mask the heterogeneous nature of CME events, which can be captured through tracking and analyzing individual CME events with multi-point measurements. On the other hand, studies of individual events do not reveal global trends. As a result, statistical studies of CME events that have multi-point measurements, are needed. This is what we perform here. 

\justify

This paper presents a comprehensive list of 47 CME events observed in longitudinal conjunction in the inner heliosphere. Previously compiled databases of CME events observed at MESSENGER, \textit{Venus Express}, STEREO, and L1 are used in association to build our conjunction database. Each conjunction event is identified based on strict directional and temporal criteria. Coronal CME counterparts to the conjunction events, the shock/discontinuity and ejecta arrival times, maximum magnetic field strength within the sheath and the ejecta, average estimated impact speeds at Mercury and Venus from the DBM, the drag parameters and standard deviations associated with various speed estimates and the measured maximum CME speed at different spacecraft as well as the average transit speeds are listed in Table-S1 and Table-S2 of the Supporting Information. We examined the variation of CME properties (propagation speed, deceleration/acceleration, magnetic field intensity, and sheath duration) with heliocentric distance, using both a statistical approach and individual analysis. We note that our statistical approach differs from previously carried out approaches because we group pairs of measurements made in conjunction rather than sampling scattered data points with no relation.  

\justify

With temporal and speed constraints, we estimated the impact speeds at Mercury and Venus using the DBM. For $\sim$94{\%} (44 out of 47) of the conjunction events in our catalog, we have plasma measurements near 1 AU. It enabled us to use the upstream solar wind speed measured near 1 AU as an input parameter of the DBM for better precision. As expected, we observed the fast CMEs to experience significantly higher speed variations on average from their propagation from the Sun to spacecraft 2 (\textit{Venus Express}, STEREO, \textit{Wind}/ACE) compared to slow CMEs (45{\%} for fast CMEs to 5{\%} for slow CMEs). Investigating where the major portion of this variation takes place in the innermost heliosphere, we found that 58-67{\%} of it happens sunward of Mercury's orbit. We also performed an analysis of the radial evolution of average transit speeds using a multilinear robust regression fitting technique (see Figure~\ref{fig:transit}). Our fit results represented a slower fall-off of the average transit speed with heliocentric distance, compared to \citeA{Winslow:2015}. It was also evident from the plot that events with the highest differences between the initial CME speeds and solar wind speeds (measured near 1 AU) had a steeper fall-off of average transit speeds compared to the rest.

\justify

We also determined the average deceleration/acceleration of CMEs between pairs of observation points in space (see Figure~\ref{fig:A}). Similar to the speed profiles, we observed the average deceleration/acceleration that the CMEs undergo during their propagation from the Sun to Mercury to be $\sim$47{\%} higher than the average between Mercury and Venus. After Venus, both the variation in propagation speeds and therefore average CME deceleration/acceleration becomes negligible. These findings lend further confirmation to past studies \cite <e.g.>[]{Gopalswamy:2001b,Reiner:2007,Winslow:2015} which suggest that variation in CME propagation speeds do occur at heliocentric distances greater than Mercury's orbit, at least to Venus's orbit.       

\justify

We examined the variation of magnetic field intensity with increasing heliocentric distance. We observed on average the maximum magnetic field strength in the ejecta to decrease by a factor of $\sim$3.4 from MESSENGER to \textit{Venus Express} and $\sim$2 from \textit{Venus Express} to near 1 AU. However, the ratio of the maximum magnetic field intensity measured in the sheath to that of the ejecta remained relatively constant throughout, which is consistent with \citeA{Janvier:2019}. We employed a multilinear robust regression fitting to investigate the dependence of the maximum magnetic field intensity measured in the sheath and the ejecta on heliocentric distance (see Figure~\ref{fig:SE}). It is important to note that our findings (scaling of the peak magnetic field strength in the ejecta with heliocentric distance) are in good agreement with previous statistical studies. Using a multilinear robust regression fitting technique, we found the following best fit power law equation for the decrease of the peak magnetic field strength in the ejecta with heliocentric distance: B$_{max,ejecta}$= $14.01^{+1.95}_{-1.71}$ $r^{-1.91\pm0.25}$. Applying the same fitting technique for MESSENGER and STEREO measurements, \citeA{Winslow:2015} reported this power law equation to be: B$_{max,ejecta}$= $12.18^{+0.75}_{-0.71}$ $r^{-1.89\pm0.14}$. From Helios observations between 0.3-1 AU, \citeA{Farrugia:2005} showed that the central axial field strength of MCs varies as $\propto r^{-1.73}$. \citeA{Leitner:2007} performed the freely expanding Lundquist flux rope fitting \cite{Farrugia:1992,Farrugia:1995} of the magnetic field for 7 MC events observed in conjunction between two or more spacecraft and found a faster decrease (B$_{max,ejecta}$$\propto r^{-2.0}$). In a study published after the initial submission of this manuscript, \citeA{Good:2019} performed a static Lundquist flux rope fitting \cite{Burlaga:1988,Lepping:1990} for 18 interplanetary flux ropes observed in conjunction between radially aligned spacecraft in the inner heliosphere. Among these, Lundquist fits for 13 events ($\chi^{2}$\textless 1.5 and fit uncertainty, $\delta$\textless 10$^{\circ}$) reproduced the observations well. They fitted the 26 ensemble values of the axial magnetic field strength (B$_{o}$, found from the flux rope fitting) against r, using an unweighted least squares linear fit to the logarithmic values of the parameters. They found the following power law equation corresponding to the radial dependence of the axial magnetic field strength: B$_{o}$=$12.5^{+3.0}_{-3.0}$ $r^{-1.76\pm0.04}$.

\justify

Again, all of these results are derived from purely statistical approaches. Taking advantage of successive observations of the same events, we found on average this decrease of the peak magnetic field strength in the ejecta to occur at a slower rate (B$_{max,ejecta}$$\propto r^{-1.75}$), similar to the value reported by \citeA{Farrugia:2005} for their nonaveraged values. The discrepancies between different statistical results can be attributed to the different fitting techniques employed. Individual analysis of pairs of measurements demonstrated this exponent decrease of the maximum magnetic field intensity in the ejecta to differ from our multilinear fit in a significant manner. This highlighted the fact that CME-to-CME variability is not well represented by fits and that many CMEs (37 out of 45 CMEs in our catalog) have a rate of decrease outside of the 95{\%} confidence interval based on our statistical approach. We highlighted this variability for both substructures in Figure~\ref{fig:SE}. For events with small longitudinal separations (\textless 10$^{\circ}$), the drop-off of the peak magnetic field strength in the sheath with heliocentric distance generally follows the power law fit equation. However, in the case for the scaling of the peak magnetic field strength in the ejecta, a higher number of events showed considerable deviations from the regression trend even for much smaller longitudinal separations (\textless 6$^{\circ}$). 

A similar trend for the axial magnetic field strength was observed by \citeA{Good:2019}. From their individual analysis of 13 well fitted events, the average fit parameters for the power law equation, B$_{o}$=$17.3^{+12.8}_{-12.8}$ $r^{-1.34\pm0.71}$ displayed large standard deviations. They attributed these considerable uncertainties being a reflection of the large spread in radial dependencies of the axial magnetic field strength displayed by individual events. Here, we find that a large event-to-event variability can still be found even if we are analyzing $\sim$3.5 times more events.

\justify

We investigated the typical duration of the sheath at different heliocentric distances, from the orbit of Mercury (0.31-0.44 AU), to that of Venus (0.72-0.73 AU) and near 1 AU. We found the sheath duration to increase with solar distance due to CME expansion. We observed the sheath duration on average to increase by a factor of $\sim$3 from MESSENGER (average sheath duration is 2.30 hours) to \textit{Venus Express} (average sheath duration is 7.07 hours) and $\sim$2 from \textit{Venus Express} to near 1 AU (average sheath duration is 11.76 hours). The sheath duration showed decent correlation with heliocentric distance when fit by a linear function (see Figure~\ref{fig:Sheath}). From this fit, we approximated the formation of the sheath to start around 0.24 AU. However, we observed the sheath duration to be fairly independent of the initial CME speed throughout the inner heliosphere with comparable sheath durations for fast and slow CMEs at Mercury, Venus, and near 1 AU. This finding provides more argument to the assumption that CME sheaths are significantly different from magnetosheaths, as suggested by \citeA{Siscoe:2008}.

\justify

We provide an example of our procedure using three spacecraft measurements for a well-studied event, also studied in \citeA{Good:2015,Good:2018}. The sheath expanded slowly from MESSENGER to STEREO-B compared to our findings. We also observed a much shorter sheath period at Venus than expected which can be due to \textit{Venus Express} measuring the nose of the ejecta rather than the flank as in the case of MESSENGER and STEREO-B. The ratio of the maximum magnetic field strength measured in the sheath to that of the ejecta remained constant except at \textit{Venus Express}, again which can be attributed to a front on encounter with the ejecta. We compared the expansion of the ejecta based on the speed profile measured at 1 AU and from the decrease of the maximum magnetic field strength in the ejecta between pairs of spacecraft. Based on the metrics of typical expansion speed \cite{Richardson:2010} and expansion parameter \cite{Gulisano:2010} at 1 AU, we found the ejecta to be expanding quickly. However, the exponent decrease of the maximum magnetic field strength from MESSENGER to STEREO-B revealed a much weaker expansion. This unexpected mismatch again confirms the argument that individual CME behavior can significantly differ from an average comparison.   

\justify

This study provides valuable insights on the evolution of CMEs in the inner heliosphere. Though statistical approaches of CME profiles are of importance, individual case studies from this catalog can capture the significant variability observed in the evolution and magnetic topology from event to event. With the catalog comprising of events observed in longitudinal separations ranging from $\sim$2 to 44$^{\circ}$, analysis of such wide range of events can provide important information regarding radial evolution and longitudinal variations in CME signatures. This catalog provides a continuous profile of CME events in terms of observations from the CME eruption at the Sun to being measured at two distinct points in space. As a result, this catalog can serve as a model validation for the prediction of CME arrival. In the coming years, missions such as Parker Solar Probe and Solar Orbiter will present more opportunities to complete the catalog of \textit{in situ} detected CMEs even closer to the Sun, but it will certainly take several years to obtain more than 40 CMEs measured at multiple spacecraft as presented here. Until then, we hope this new catalog of CMEs observed in conjunction between inner heliosphere spacecraft will prove valuable to the community.

\begin{acknowledgments}

\justify

The authors acknowledge the use of MESSENGER data available at Planetary Data System (\url{https://pds.nasa.gov/}), \textit{Venus Express} data available at European Space Agency's Planetary Science Archive (\url{https://www.cosmos.esa.int/web/psa/venus-express}) and NASA/GSFC's Space Physics Data Facility's CDAWeb service for STEREO, \textit{Wind}, and ACE data available at CDAWeb (\url{https:cdaweb.gsfc.nasa.gov/index.html/}). The authors acknowledge Christian M{\"o}stl and Miho Janvier for providing \textit{Venus Express} data and helping in the coordinate transformation. We also thank the two anonymous
reviewers for their contribution in improving the manuscript. Associated datasets for this research are available at \url{http://doi.org/10.5281/zenodo.3592426}. T.~M.~S. was supported by NASA grant 80NSSC17K0556 and NSF grant AGS1435785. N.~L. acknowledges support from NASA grants NNX15AB87G and 80NSSC17K0556 and NSF grant AGS1435785. R.~M.~W. acknowledges support from NSF grant AGS1622352, and NASA grants NNX15AW31G and 80NSSC19K0914.

\end{acknowledgments}

\bibliography{Salman}


\end{document}